\documentclass[11pt, a4paper]{article}
\pdfoutput=1
\usepackage{jcappub}

\usepackage{booktabs}
\usepackage{multirow}
\usepackage{amssymb}
\usepackage[export]{adjustbox}
\usepackage{floatrow}
\newfloatcommand{capbtabbox}{table}[][3.5in]
\newfloatcommand{capbfigbox}{figure}[][2.3in]
\usepackage{blindtext}
\usepackage{amsmath}
\usepackage{booktabs}
\usepackage{slashed}
\usepackage{cancel}
\usepackage{aas_macros}
\usepackage{centernot}

\newcommand{\ie}{i.e.~}

\def\lsim{\mathrel{\raise.3ex\hbox{$<$\kern-.75em\lower1ex\hbox{$\sim$}}}}
\def\gsim{\mathrel{\raise.3ex\hbox{$>$\kern-.75em\lower1ex\hbox{$\sim$}}}}

\usepackage{appendix}
\usepackage{array}
\newcolumntype{x}[1]{>{\centering\arraybackslash}p{#1}}
\usepackage{bm}
\usepackage{color}
\usepackage{graphicx}
\usepackage{cancel}
\usepackage{amsfonts}
\usepackage{amssymb}
\usepackage{amsmath}
\usepackage{calc}
\usepackage{dcolumn}
\usepackage{mathrsfs}
\usepackage{mathtools}
\usepackage{soul}

\usepackage{braket}
\usepackage[normalem]{ulem}

\newcommand{\Fig}[1]{Fig.~\ref{#1}}

\newcommand{\beq}{\begin{equation}}
\newcommand{\eeq}{\end{equation}}

\definecolor{rossoCP3}{cmyk}{0,.88,.77,.40}
\definecolor{verdeCP3}{rgb}{0.09765625, 0.57421875, 0.1015625}
\definecolor{bluCP3}{rgb}{0, 0.23, 0.67}

\ifx\pdfoutput\undefined
\usepackage[dvips,bookmarks]{hyperref}    
\else
\usepackage{hyperref}    
\fi
\hypersetup{colorlinks, bookmarksopen, bookmarksnumbered,
citecolor=verdeCP3, linkcolor=bluCP3, pdfstartview=FitH, urlcolor=rossoCP3}

\begin{document}


\hspace*{80mm}{\tt IFIC/17-38, FERMILAB-PUB-17-383-A}
\vskip 0.2in


\title{Hidden Sector Dark Matter and the Galactic Center Gamma-Ray Excess: A Closer Look}

\author[a]{Miguel Escudero,}\note{ORCID: http://orcid.org/0000-0002-4487-8742}
\emailAdd{miguel.escudero@ific.uv.es}
\author[a,b]{Samuel J.~Witte}\note{ORCID: http://orcid.org/0000-0003-4649-3085}
\emailAdd{switte@physics.ucla.edu}
\author[c,d,e]{and Dan Hooper}\note{ORCID: http://orcid.org/0000-0001-8837-4127}
\emailAdd{dhooper@fnal.gov}

\affiliation[a]{Instituto de F\'{\i}sica Corpuscular (IFIC)$,$ CSIC-Universitat de Val\`encia$,$ Apartado de Correos 22085$,$ E-46071 Valencia$,$ Spain}
\affiliation[b]{University of California, Los Angeles, Department of Physics and Astronomy, Los Angeles, CA 90095}
\affiliation[c]{Fermi National Accelerator Laboratory, Center for Particle
Astrophysics, Batavia, IL 60510}
\affiliation[d]{University of Chicago, Department of Astronomy and Astrophysics, Chicago, IL 60637}
\affiliation[e]{University of Chicago, Kavli Institute for Cosmological Physics, Chicago, IL 60637}

\abstract{Stringent constraints from direct detection experiments and the Large Hadron Collider motivate us to consider models in which the dark matter does not directly couple to the Standard Model, but that instead annihilates into hidden sector particles which ultimately decay through small couplings to the Standard Model. We calculate the gamma-ray emission generated within the context of several such hidden sector models, including those in which the hidden sector couples to the Standard Model through the vector portal (kinetic mixing with Standard Model hypercharge), through the Higgs portal (mixing with the Standard Model Higgs boson), or both. In each case, we identify broad regions of parameter space in which the observed spectrum and intensity of the Galactic Center gamma-ray excess can easily be accommodated, while providing an acceptable thermal relic abundance and remaining consistent with all current constraints. We also point out that cosmic-ray antiproton measurements could potentially discriminate some hidden sector models from more conventional dark matter scenarios.}

\maketitle


\newpage

\section{Introduction}

A number of groups have reported the presence of a significant excess of GeV-scale gamma rays from the region surrounding the Galactic Center~\cite{Goodenough:2009gk,Hooper:2010mq,Hooper:2011ti,Abazajian:2012pn,Gordon:2013vta,Hooper:2013rwa,Daylan:2014rsa,Calore:2014xka,TheFermi-LAT:2015kwa,Karwin:2016tsw,TheFermi-LAT:2017vmf}, with spectral and morphological characteristics that are broadly consistent with that predicted from annihilating dark matter particles. Although this signal's possible connection with dark matter has received a great deal of attention (see, for example, Refs.~\cite{Ipek:2014gua,Izaguirre:2014vva,Agrawal:2014una,Berlin:2014tja,Alves:2014yha,Boehm:2014hva,Huang:2014cla,Cerdeno:2014cda,Okada:2013bna,Freese:2015ysa,Fonseca:2015rwa,Bertone:2015tza,Cline:2015qha,Berlin:2015wwa,Caron:2015wda,Cerdeno:2015ega,Liu:2014cma,Hooper:2014fda,Arcadi:2014lta,Cahill-Rowley:2014ora,Ko:2014loa,McDermott:2014rqa,Abdullah:2014lla,Martin:2014sxa,Berlin:2014pya,Hooper:2012cw,Cirelli:2016rnw}), astrophysical origins of this emission have also been extensively discussed. In particular, scenarios have been proposed in which the GeV excess is generated by a large population of unresolved millisecond pulsars~\cite{Cholis:2014lta,Petrovic:2014xra,Brandt:2015ula,Lee:2015fea,Hooper:2015jlu,Bartels:2015aea,Hooper:2016rap}, or by a series of recent cosmic-ray outbursts~\cite{Cholis:2015dea,Petrovic:2014uda,Carlson:2014cwa}. Outburst scenarios, however, require a significant degree of tuning in their parameters~\cite{Cholis:2015dea}, and pulsars can generate this signal only if the population of these objects in the Inner Galaxy is different from those observed in globular clusters or in the field of the Milky Way~\cite{Haggard:2017lyq,Hooper:2016rap,Hooper:2015jlu,Cholis:2014lta}.

Dark matter scenarios capable of accounting for the observed gamma-ray excess are also quite strongly constrained. In particular, the null results of direct detection experiments~\cite{Aprile:2017iyp,pandax,Akerib:2016vxi}, as well as the LHC (Large Hadron Collider) and other collider experiments, exclude many models in which the dark matter is an electroweak-scale thermal relic. Although there exist models in which the dark matter could generate the gamma-ray excess without violating these stringent constraints (featuring pseudoscalar mediators, or near resonance spin-1 mediators, for example)~\cite{Escudero:2016kpw}, these results motivate us to consider models in which the dark matter does not couple directly to the particle content of the Standard Model, but instead produce other hidden sector particles in their annihilations, which decay through very small couplings to the Standard Model. Such hidden sector dark matter models have been previously explored, including within the context of the Galactic Center excess~\cite{Abdullah:2014lla,Berlin:2014pya,Martin:2014sxa,Hooper:2012cw}.

In this paper, we revisit the possibility that the Galactic Center gamma-ray excess may be generated by the annihilations of hidden sector dark matter particles. In the following section, we describe three such models, and calculate in each the dark matter's thermal relic abundance, low-velocity annihilation cross section, and elastic scattering cross section with nuclei. We then go on to calculate the gamma-ray spectrum that is generated through dark matter annihilations in each model, and compare these results to the observed spectral shape and intensity of the Galactic Center excess. In each case we find broad regions of parameter space that can provide a good fit to the observed characteristics of the gamma-ray excess. We also discuss additional constraints on the vector and Higgs portal scenarios, and consider how hidden sector dark matter models can be further tested and probed in the future, including through cosmic-ray antiproton measurements.

\section{Hidden Sector Dark Matter}

Hidden sector models fall into broad classifications, depending on the interactions which connect the hidden sector to the particle content of the Standard Model~\cite{Pospelov:2007mp,ArkaniHamed:2008qn}. Particularly attractive are the scenarios known as the vector portal~\cite{Holdom:1985ag}, the Higgs portal~\cite{Silveira:1985rk,Patt:2006fw,McDonald:1993ex,Burgess:2000yq}, and the neutrino portal~\cite{Minkowski:1977sc,GellMann:1980vs,Yanagida:1979as,Mohapatra:1979ia}. In this study, we focus on our attention on the first two of these possibilities, which are described in the following two subsections (for recent studies of dark matter phenomenology in neutrino portal models, see Refs.~\cite{Falkowski:2011xh,Tang:2015coo,Macias:2015cna,Gonzalez-Macias:2016vxy,Escudero:2016tzx,Escudero:2016ksa,Batell:2017rol,Campos:2017odj}). We then describe three dark matter models which incorporate these portals between the hidden sector and the Standard Model.

\subsection{The Vector Portal}\label{subsec:kinetic}

\begin{figure}[t]
\centering
\includegraphics[scale=0.5]{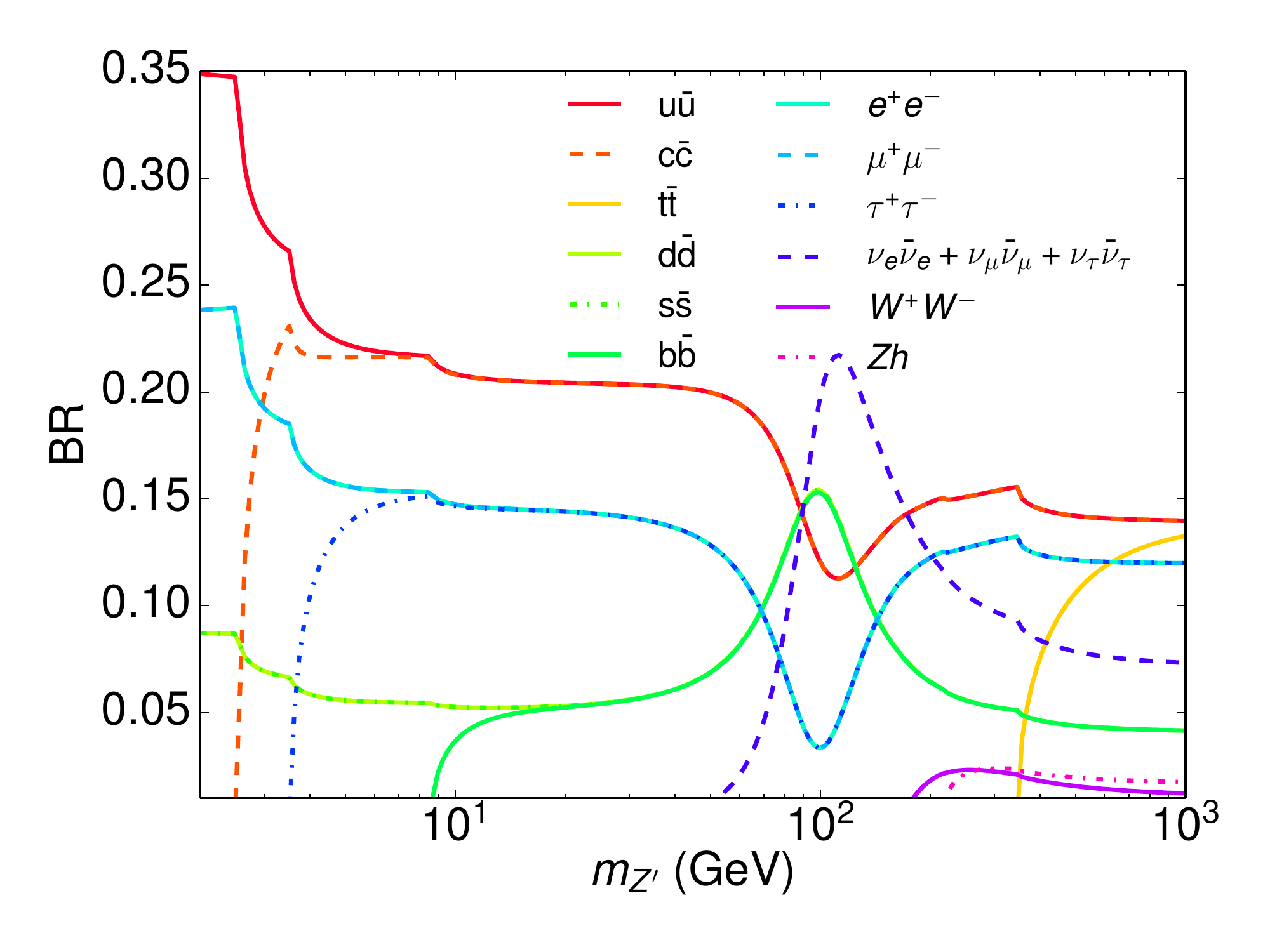}
\caption{The branching ratios of a hidden sector $Z'$ which couples to the Standard Model through kinetic mixing with Standard Model hypercharge.}\label{fig:BRZprime}
\end{figure}

We begin by considering a hidden sector which contains a broken $U_D(1)$ symmetry, resulting in a massive gauge boson, $Z'$. We also assume that no scalar fields are charged simultaneously under both the Standard Model and hidden sector gauge groups. The kinetic terms in the hidden sector Lagrangian are given by: 
\begin{equation}
	\mathcal{L}_\text{kinetic} \supset  -\frac{1}{4}\hat{B}_{\mu\nu}\hat{B}^{\,\mu\nu}  -\frac{1}{4}\hat{F'}_{\mu\nu}\hat{F'}^{\,\mu\nu} - \frac{\epsilon}{2}\hat{B}_{\mu\nu}\hat{F'}^{\,\mu\nu} \, ,
\end{equation}
where $\hat{B}_{\mu \nu}$ and $\hat{F'}_{\mu\nu}$ are the stress-energy tensors of $U_Y(1)$ and $U_D(1)$, respectively. After the diagonalization of the kinetic terms, and subsequently the mass terms, the relevant interactions between the hidden sector $Z'$ and the Standard Model fermions are described by the following (assuming $\epsilon \ll 1$)~\cite{Babu:1997st,Cline:2014dwa,Gopalakrishna:2008dv}:
\begin{eqnarray}\label{eq:ZprimetoSM}
	\mathcal{L}_{Z',\, \text{fermions}} &=& Z'^\mu\sum_f \bar{f} \gamma_\mu (g_v^f + g_a^f \gamma_5) f \,, 
\end{eqnarray}
where the vector and axial couplings are given as follows:
\begin{eqnarray}
\label{couplingseq}
g_v^f &=& \frac{\epsilon}{(m^2_Z-m^2_{Z'})} \, \bigg[m^2_{Z'} \, g_Y \, \frac{(Y_{f_R}+Y_{f_L})}{2} -m^2_Z \, g \, \sin \theta_W  \cos \theta_W \, Q_f   \bigg], \nonumber \\
g_a^f &=& \frac{\epsilon}{(m^2_Z-m^2_{Z'})} \, \bigg[m^2_{Z'} \, g_Y \, \frac{(Y_{f_R}-Y_{f_L})}{2}   \bigg] \,.
\end{eqnarray}
Here, $Q_f$ and $Y_{f_{R,L}}$ denote electric charge and hypercharge, respectively, $\theta_W$ the weak mixing angle, and $g_Y$ and $g$ are the Standard Model $U(1)$ and $SU(2)$ gauge couplings, respectively.

The branching ratios for the $Z'$ are shown in Fig.~\ref{fig:BRZprime}. For values of $m_{Z'}$ which are either much greater than or much less than $m_Z$, these decays are dominated by final states consisting of charged leptons and up-type quarks, due to the large values of these fermions' hypercharge. For $m_{Z'} \sim m_Z$, however, a cancellation occurs, leading to a large branching fraction to neutrinos and down-type quarks.

If $\epsilon$ is not too small, interactions of the $Z'$ can maintain kinetic equilibrium between the hidden sector and the Standard Model in the early universe. In particular, processes of the type $Z' f \leftrightarrow \gamma f$ will exceed the rate of Hubble expansion at a temperature $T$ if the following condition is satisfied: $\epsilon \gsim 3\times 10^{-8} \times (T/{\rm GeV})^{1/2} \, (g_{\star}/75)^{1/4}$, where $g_{\star}$ is the effective number of degrees-of-freedom at temperature $T$ (see, for example, Appendix 7 of Ref.~\cite{Berlin:2016vnh}).

\subsection{The Higgs Portal}\label{subsec:Higgsportal}

We next consider interactions between the hidden sector and the Standard Model which are generated through mixing with the Higgs boson. We introduce a complex scalar, $\phi$, which transforms as a singlet under the Standard Model gauge symmetries and that is charged under a new local $U_D(1)$ symmetry. Including all renormalizable interactions, this symmetry leads to the following scalar potential:
\begin{eqnarray} \label{eq:scalarpot}
	V &=& -\mu_H^2(H^\dagger H) + \lambda_H (H^\dagger H)^2 - \mu_\phi^2 \phi^\dagger \phi + \lambda_\phi (\phi^\dagger \phi)^2 + \lambda_{H\phi} (H^\dagger H) \, (\phi^\dagger \phi) \, ,
\end{eqnarray}
where $H$ is the Standard Model Higgs doublet. After both electroweak and dark symmetry breaking, both scalars develop vacuum expectation values, so that in the unitary gauge
\begin{equation}
H = \left( \begin{array}{c}
0\\
\frac{v_H + \tilde h }{\sqrt{2}}
\end{array}
\right)  \ , \qquad \phi = \frac{v_\phi + \tilde \rho}{\sqrt{2}} \, .
\\
\end{equation}
The scalar sector then contains two CP even massive real scalars, $\tilde h$ and $\tilde \rho$, which mix. Upon diagonalization of the mass matrix, this leads to the mass eigenstates $h$ and $\rho$. The state $h$ is identified as the Standard Model Higgs boson with a mass of $m_h  \approx 125$ GeV. The mixing angle between these two states is given by:
\begin{eqnarray}
	\tan 2\theta = \frac{\lambda_{H\phi}  \, v_H \, v_\phi }{\lambda_\phi \, v_\phi^2 - \lambda_H \, v_H^2} \, .
\end{eqnarray}
The remaining couplings can be written in terms of the physical masses and this mixing angle:
\begin{eqnarray}
	\lambda_H = \frac{m_h^2 \cos^2 \theta + m_\rho^2\sin^2 \theta}{2 v_H^2}\, ,& \qquad & \lambda_\phi = \frac{m_h^2 	\sin^2 \theta + m_\rho^2 \cos^2\theta }{2 v_\phi^2} \, , \qquad \lambda_{H\phi} = \frac{( m_\rho^2 - m_h^2) \sin 2 	\theta }{2 v_H v_\phi} \,, \nonumber \\
	\mu_H^2 = \lambda_H  v_H^2 + \frac 1 2 \lambda_{H \phi} v_\phi^2 \ ,  &\qquad& \mu_\phi^2 = \lambda_\phi  v_\phi^2 + \frac 1 2 \lambda_{H \phi} v_H^2 \ .
\end{eqnarray}
We have implicitly assumed that the CP-odd state contained in $\phi$ corresponds to the longitudinal mode of the hidden sector $Z'$, so that it gets its mass through the Higgs mechanism, namely $m_{Z'} = 2 g_D v_\phi $, where $g_D$ is the coupling strength of the $U_D(1)$ symmetry and we have assumed that the $\phi$ field carries two units of $U_D(1)$ charge. Due to this mixing, the couplings of the Higgs boson to Standard Model particles are rescaled by a factor of $\cos \theta$, such that the total width is given by $\Gamma(h \to \text{SM}) = \cos^2\theta \, \Gamma_\text{SM}$, where $\Gamma_\text{SM} = 4.07 \, \text{MeV}$. In addition, if $m_\rho < m_h/2$, the Higgs boson will be able to decay to $\rho \rho$, with a width that is given as follows:
\begin{eqnarray}	
 \Gamma(h \rightarrow \rho \rho) = \frac{(m_h^2 + 2 m_\rho^2)^2}{128 \pi m_h^2 v_H^2 v_\phi^2} (m_h^2 - 4 m_\rho^2)^{1/2} \, (v_H \cos\theta - v_\phi \sin\theta )^2  \sin^2 2\theta \, .
\end{eqnarray}

\begin{figure}[t]
\centering
\includegraphics[scale=0.5]{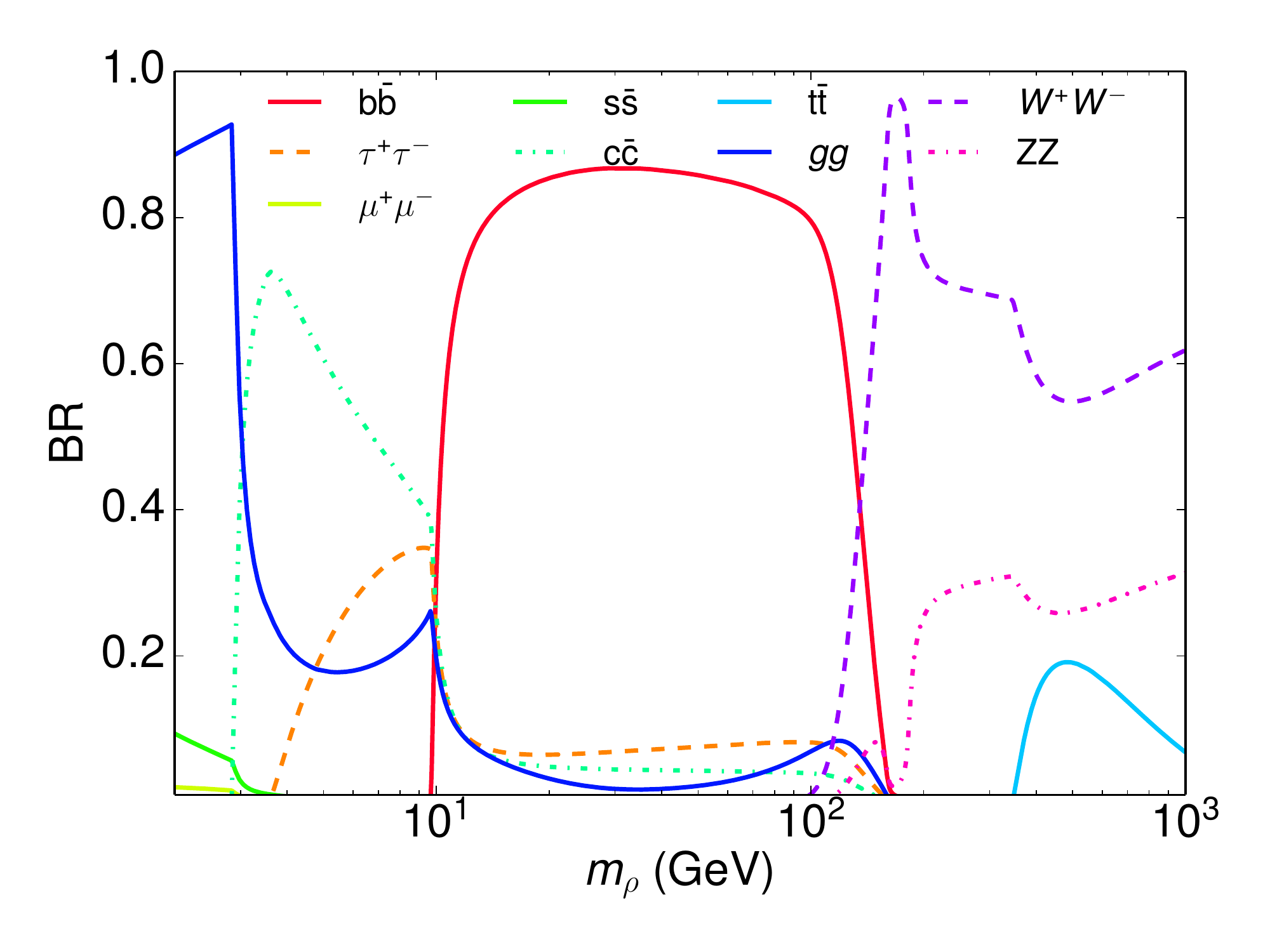}
\caption{The branching ratios of a hidden sector scalar, $\rho$, which couples to the Standard Model through mixing with the Higgs boson.}\label{fig:BRHdark}
\end{figure}

The branching ratios of this hidden sector scalar are shown in Fig.~\ref{fig:BRHdark}, as computed using HDECAY~\cite{Djouadi:1997yw}. As expected, these decays are dominated by final states which include Standard Model gauge bosons and heavy fermions.

If $\lambda_{H\phi}$ is not too small, kinetic equilibrium will be maintained between the hidden sector and the Standard Model in the early universe. By comparing the rate for $\phi \phi \leftrightarrow HH$ to that of Hubble expansion, we find that kinetic equilibrium is maintained between these sectors at $T\sim {\rm GeV}$ temperatures as long as $\lambda_{H\phi} \gsim 10^{-7}$, corresponding to $\sin^2 \theta \gtrsim 10^{-13} \left(v_\phi/300 \, \text{GeV}  \right)^2$.

\subsection{Dirac Dark Matter and the Vector Portal}\label{subsec:Dirac}

First, we will consider dark matter in the form of a Dirac Fermion, $\Psi$, which is charged under a broken $U(1)_\text{D}$ and thus couples to a hidden sector $Z'$. In particular, we will consider the case in which the $Z'$ gets its mass via the Stueckelberg mechanism~\cite{Kors:2004dx} and mixes with Standard Model hypercharge as described in Sec.~\ref{subsec:kinetic}. A Dirac mass term for $\Psi$ is allowed in the Lagrangian, and the dark matter's stability is ensured by a residual $Z_2$ symmetry. Among other studies, this model has been considered previously in Ref.~\cite{Cline:2014dwa} within a similar context. 

The relevant Lagrangian of this model is given by:
\begin{equation}
	\mathcal{L} \supset  \bar{\Psi}(i\slashed{\partial} -m_\Psi )\Psi + g_D Z'^\mu \bar{\Psi}\gamma_\mu \Psi \, .\label{eq:DiracLag}
\end{equation}
The early universe phenomenology of the hidden sector is dominated by the process $\bar{\Psi} \Psi \to Z' Z'$, with a cross section that is given by:
\begin{eqnarray}
\sigma_{\Psi \Psi \to Z' Z'} &=& \frac{g_D^4}{8 \pi  s \left(s-4  m_\Psi^2\right) \left(2 m_{Z'}^2-s\right)} \nonumber \\
&\times& \left[ \frac{\left(s-2 m_{Z'}^2\right) \sqrt{\left(s-4  m_\Psi^2\right) \left(s-4 m_{Z'}^2\right)} \left(4  m_\Psi^4+ m_\Psi^2 s+2 m_{Z'}^4\right)}{ m_\Psi^2 \left(s-4 m_{Z'}^2\right)+m_{Z'}^4}\right.  \\ 
&+& \left. \left(8  m_\Psi^4+ m_\Psi^2 \left(8 m_{Z'}^2-4 s\right)-4 m_{Z'}^4-s^2\right) \ln \left(\frac{-\sqrt{\left(s-4  m_\Psi^2\right) \left(s-4 m_{Z'}^2\right)}+2 m_{Z'}^2-s}{\sqrt{\left(s-4  m_\Psi^2\right) \left(s-4 m_{Z'}^2\right)}+2 m_{Z'}^2-s}\right)  \right]. \nonumber
\end{eqnarray}
In the low-velocity limit (relevant for indirect detection), this reduces to:
 \begin{eqnarray}
      \sigma_{\Psi \bar{\Psi} \to Z' Z'}v = \frac{g_D^4}{ 4 \pi  m_\Psi \left(m_{Z'}^2-2 m_\Psi^2\right)^2}  \left[m_\Psi^2-m_{Z'}^2\right]^{3/2} + \mathcal{O}(v^2)\,.
\end{eqnarray}

For the purposes of direct detection, the dominant process is spin-independent scattering, with the following cross section per nucleon:
\begin{eqnarray}
\sigma^\text{Dirac}_\text{nucleon} &=& \frac{\mu_{\Psi N}^2 g_D^2}{\pi m_{Z'}^4} \left[g_v^u(1+Z/A)+ g_v^d(2-Z/A) \right]^2 \, ,
\end{eqnarray}
where $g_v^{u,d}$ are as defined in Eq.~\ref{couplingseq}, $\mu_{\Psi N}$ is the reduced mass of the dark matter-nucleon system, and $A$ and $Z$ are the atomic mass and number of the target nucleus, respectively.

\begin{figure}[t]
\begin{center}
\begin{tabular}{cc}
\includegraphics[width=0.495\textwidth]{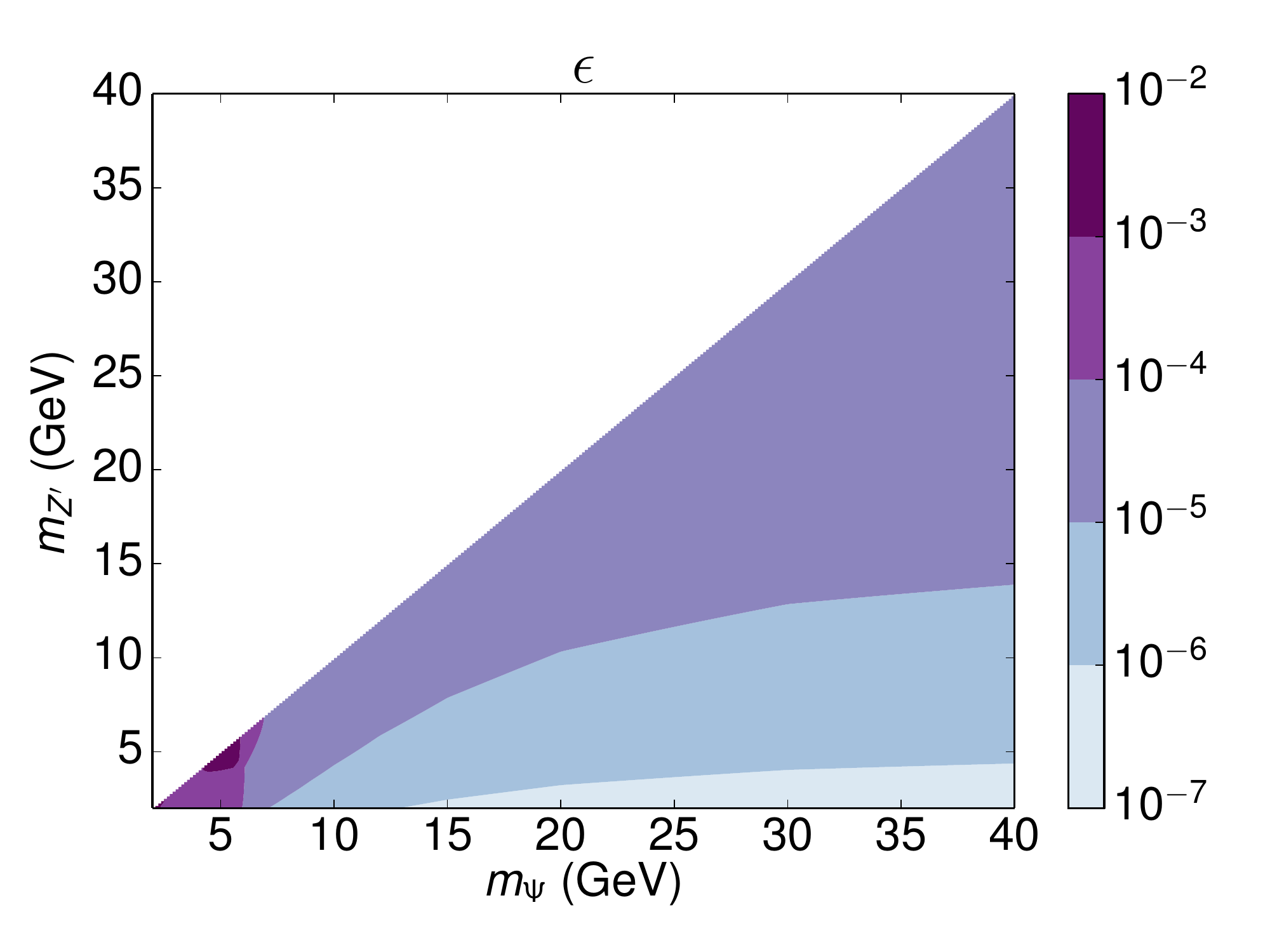} 
\includegraphics[width=0.495\textwidth]{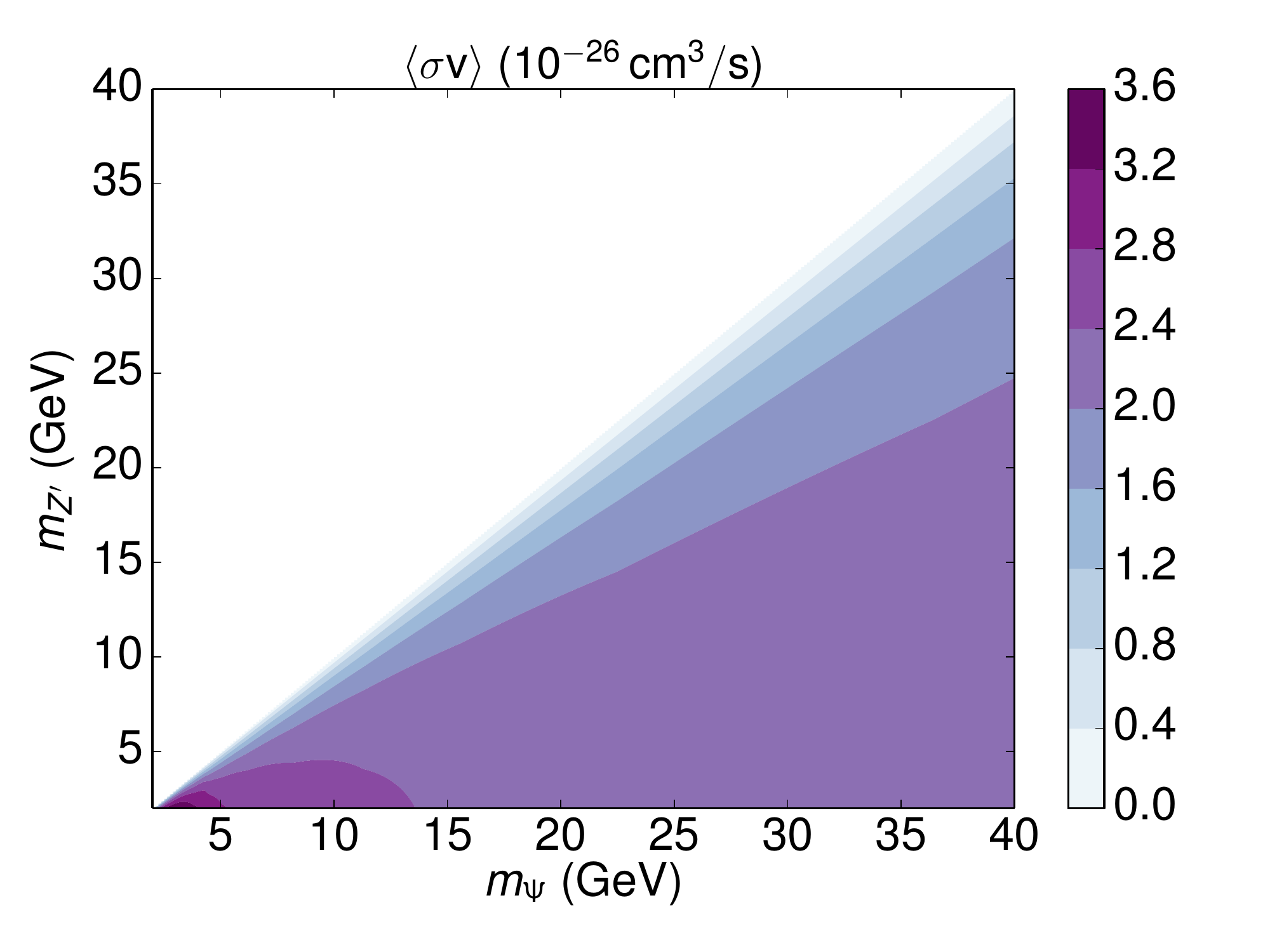} 
\end{tabular}
\caption{A summary of the phenomenology in a model with a Dirac fermion dark matter candidate, $\Psi$, which annihilates to a $Z'$ pair, which then decay through kinetic mixing with Standard Model hypercharge. Throughout each frame, we have chosen the value of the hidden sector gauge coupling in order to obtain a thermal relic abundance equal to the measured cosmological dark matter density. In the left frame, we plot the maximum value of the kinetic mixing parameter, $\epsilon$, as derived from direct detection constraints~\cite{Aprile:2017iyp,Angloher:2015ewa,Armengaud:2012pfa}. In the right frame, we plot the annihilation cross section (in units of $10^{-26}$ cm$^3/$s) evaluated at a velocity of $v=10^{-3} c$, as appropriate for indirect searches.}\label{fig:Dirac_res}
\end{center}
\end{figure}

In Fig.~\ref{fig:Dirac_res}, we summarize some aspects of the dark matter phenomenology in this model. Throughout each frame, we have chosen the value of the coupling $g_D$ in order to obtain a thermal relic abundance equal to the measured cosmological dark matter density. Throughout this study, we calculate the relic abundance following the method described in Refs.~\cite{Cline:2013gha,Cline:2014dwa} (see also Refs.~\cite{Steigman:2012nb,Gondolo:1990dk,Griest:1990kh}). In doing so, we assume that $\epsilon$ is large enough to maintain kinetic equilibrium between the hidden sector and the Standard Model at the time of dark matter freeze-out (see Sec.~\ref{subsec:kinetic}). If this condition is not satisfied, the elastic scattering cross section with nuclei and the low-velocity annihilation cross section could be either larger or smaller than those shown here~\cite{Berlin:2016vnh,Berlin:2016gtr}. 

In the left frame of Fig.~\ref{fig:Dirac_res}, we plot the maximum value of $\epsilon$, as derived from direct detection constraints~\cite{Aprile:2017iyp,Angloher:2015ewa,Armengaud:2012pfa}. In the right frame, we plot the annihilation cross section evaluated at a velocity of $v=10^{-3} c$, as appropriate for indirect searches. Note that for $m_{\Psi} \sim m_{Z'}$, the low-velocity cross section is reduced, due to differing velocity distributions in the early and contemporary universe.

\subsection{Vector Dark Matter and the Higgs Portal}\label{subsec:Vector}

Fermionic dark matter can annihilate to a pair of spin-$0$ particles efficiently in the low-velocity limit only through a product of both scalar and pseudoscalar couplings. Alternatively, we can also consider either scalar or vector dark matter, either of which can annihilate efficiently to a pair of scalars at low velocities. Here we focus on the case of a model that includes a vector dark matter candidate, $X$, and a scalar, $\phi$, with charge assignment 2~\cite{Ko:2014gha,Ko:2015ioa}. We further impose a $Z_2$ symmetry to stabilize the dark matter, and which also prohibits the possibility of any kinetic mixing. The relevant Lagrangian contains the following terms (in addition to those corresponding to the scalar potential of Eq.~\ref{eq:scalarpot}):
\begin{equation}
	\mathcal{L} \supset  -\frac{1}{4}{X}_{\mu\nu}{X}^{\,\mu\nu} + \left( D^\mu \phi\right)^\dagger \left(D_\mu \phi \right)  \equiv -\frac{1}{4}{X}_{\mu\nu}{X}^{\,\mu\nu} + \frac{1}{2} \partial_\mu  \tilde{\rho} \partial^\mu  \tilde{\rho} + \frac{1}{2} m_X^2 X_\mu X^\mu \left( 1+ 2 \frac{\tilde{\rho}}{v_\phi} + \frac{\tilde{\rho}^2}{v_\phi^2}   \right) \, ,\label{eq:Vector_Hdark}
\end{equation}
where $m_X = 2 g_D v_\phi$. Note that in this model the dark matter candidate X corresponds to the $U(1)_D$ gauge boson, although due to the $Z_2$ symmetry $\epsilon = 0$.

In this model, the dark matter annihilation cross section in the limit $\sin^2\theta \rightarrow 0$ is given as follows:
\begin{eqnarray}
\sigma_{X X \to \rho \rho } &=&  \frac{1}{288 \pi  s v_\phi^4 \left(s-4 m_X^2\right)} \\
&\times& \left[  \frac{2\ln \left(\frac{ 2 m_\rho^2+s\left(\sqrt{\beta_X \beta_\rho }-1\right)  }{ 2 m_\rho^2-s(\sqrt{\beta_X \beta_\rho } +1)}\right)^2}{2 m_\rho^4-3 m_\rho^2 s+s^2} \left\{48 m_X^8 \left(m_\rho^2-s\right)-8 m_X^6 \left(16 m_\rho^4-4 m_\rho^2 s-3 s^2\right)\right. \right. \nonumber \\
&+&\left.\left.  4 m_X^4 \left(10 m_\rho^6+2 m_\rho^4 s-3 m_\rho^2 s^2\right) + 2 m_X^2 m_\rho^2 s \left(4 m_\rho^4-5 m_\rho^2 s+s^2\right)+m_\rho^6 \left(-3 m_\rho^4-m_\rho^2 s+s^2\right)\right \}    \right. \nonumber \\
&+&\left.\frac{\sqrt{\beta_X \beta_\rho } s}{6 \left(m_\rho^2-s\right)^2} \left\{ 72 m_X^4 \left(2 m_\rho^2+s\right)^2+4 m_X^2 \left(20 m_\rho^6-75 m_\rho^4 s+s^3\right)+24 m_\rho^8+28 m_\rho^6 s \right. \right. \nonumber \\
&-& \left. \left. 3 m_\rho^4 s^2+6 m_\rho^2 s^3-s^4\right\} + \frac{1}{6} \beta_X \beta_\rho ^{3/2} s^3 \right. \nonumber \\
&+& \left.  \frac{2 \sqrt{\beta_X \beta_\rho } s \left(48 m_X^8-32 m_X^6 m_\rho^2+4 m_X^4 \left(6 m_\rho^4-4 m_\rho^2 s+s^2\right)+4 m_X^2 m_\rho^4 \left(s-2 m_\rho^2\right)+m_\rho^8\right)}{m_X^2 \left(s-4 m_\rho^2\right)+m_\rho^4}  \right] \, , \nonumber
 \end{eqnarray}
where $\beta_{\rho,X} = \sqrt{1-4 m_{\rho,X}^2/s}$. In the low-velocity limit, this reduces to:
\begin{eqnarray}
 \sigma_{X X \to \rho \rho }v = \frac{m_X^2 \sqrt{1-m_\rho^2/m_X^2}}{36 \pi  v_\phi^4 } \, \frac{\left(176 m_X^8-320 m_X^6 m_\rho^2+240 m_X^4 m_\rho^4-80 m_X^2 m_\rho^6+11 m_\rho^8\right)}{\left(8 m_X^4-6 m_X^2 m_\rho^2+m_\rho^4\right)^2} \, .\nonumber \\
\end{eqnarray}

Interactions thorough the Higgs portal (see Sec.~\ref{subsec:Higgsportal}) lead to the following spin-independent scattering cross section with nuclei: 
\begin{eqnarray}
\label{elastichiggsportal}
    	\sigma^\text{Vector}_{\rm nucleon} = \frac{f_N^2}{4\pi}\frac{\mu_{X N}^2 m_N^2}{m_X^2} \left[\frac{m_X^2 }{v_\phi v_H} \frac{\sin 2\theta}{2} \left( \frac{1}{m_\rho^2} - \frac{1}{m_h^2}  \right)
 \right]^2,
  \end{eqnarray}  
where $f_N\simeq 0.3$ and $m_N$ is the nucleon mass.

\begin{figure}[t]
\centering
\includegraphics[width=0.495\textwidth]{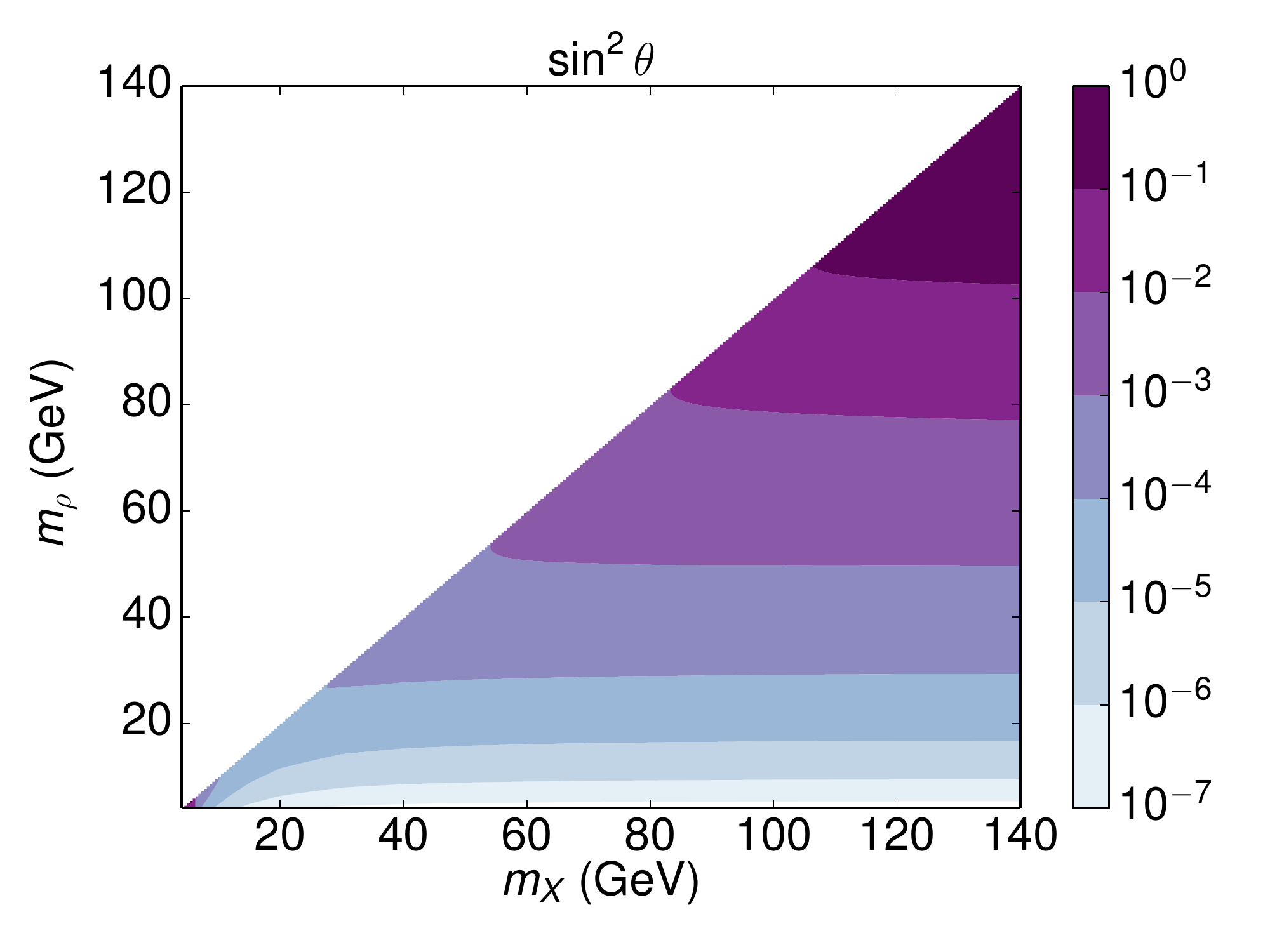}  
\includegraphics[width=0.495\textwidth]{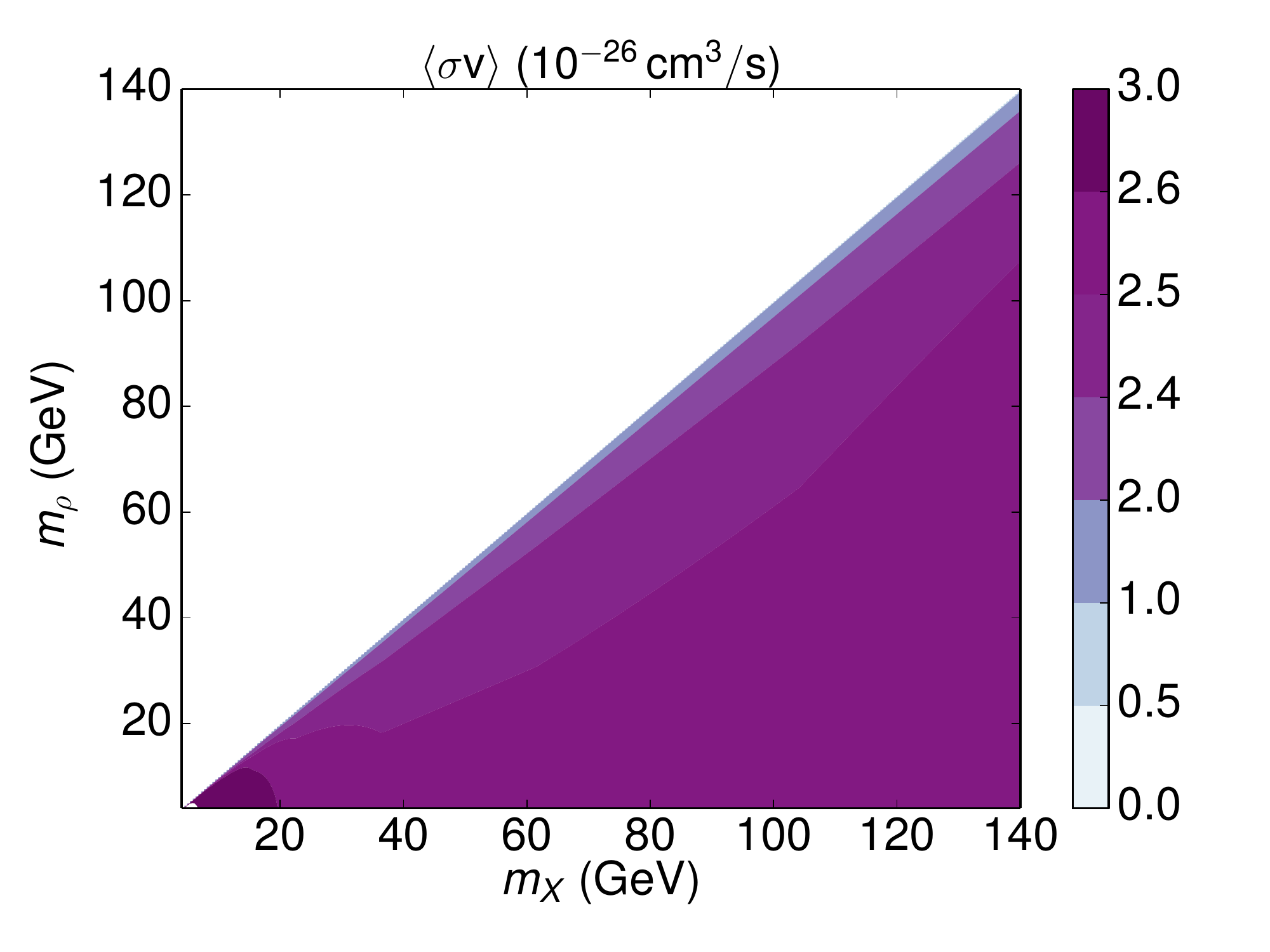}  
\caption{A summary of the phenomenology in a model with a vector dark matter candidate, $X$, which annihilates to a pair of scalars, $\rho$, which decay through mixing with the Standard Model Higgs boson. Throughout each frame, we have chosen the value of the hidden sector vacuum expectation value in order to obtain a thermal relic abundance equal to the measured cosmological dark matter density. In the left frame, we plot the maximum value of the Higgs mixing parameter, $\sin^2 \theta$, as derived from direct detection constraints~\cite{Aprile:2017iyp,Angloher:2015ewa,Armengaud:2012pfa}. In the right frame, we plot the annihilation cross section (in units of $10^{-26}$ cm$^3/$s) evaluated at a velocity of $v=10^{-3} c$, as appropriate for indirect searches.}\label{fig:Vector_res}
\end{figure}

After fixing the value of $v_{\phi}$ to obtain the desired relic abundance, we calculate both the elastic scattering cross section with nuclei and the low-velocity annihilation cross section. These results are shown in Fig.~\ref{fig:Vector_res}.

We note that there exists significant parameter space within this model in which the mixing between the $\rho$ and the Standard Model Higgs boson is quite significant, especially around $m_{\rho} \sim 125$ GeV, where there is a cancellation in the dark matter's elastic scattering cross section with nuclei (see Eq.~\ref{elastichiggsportal}). In this region of parameter space, the process $X X \to  h \rho$ can account for up to $\sim20\%$ of dark matter annihilations, and $X X \to h h$ can account for at most $\sim 4\%$, after applying constraints from colliders, \ie $\sin^2\theta \lesssim 0.1$ for all values of $m_\rho$ (see Sec.~\ref{otherconstraints}). We note that the gamma-ray spectrum that results from these channels are fairly similar to that from $XX \to \rho \rho$ in this mass range. Furthermore, the dark matter could also annihilate directly into Standard Model fermions thorough the Higgs resonance, although this channel is relevant only in a very narrow region of parameter space near $m_X \simeq m_h/2$~\cite{Escudero:2016gzx}, and we do not consider this possibility further.

In these calculations, we have assumed that $\sin^2 \theta$ is large enough to maintain kinetic equilibrium between the hidden sector and the Standard Model at the time of dark matter freeze-out (see Sec.~\ref{subsec:Higgsportal}). If this is not the case, the elastic scattering cross section with nuclei and the low-velocity annihilation cross section could be either larger or smaller than those shown in Fig.~\ref{fig:Vector_res}~\cite{Berlin:2016vnh,Berlin:2016gtr}.

\subsection{Majorana Dark Matter and Combined Vector and Higgs Portals}\label{subsec:Majorana}

Lastly, we consider Majorana dark matter in the presence of two additional hidden sector states, namely a $Z'$ and a scalar, $\rho$. Under a $U_D(1)$ symmetry, we assign charges of 2 for the $\rho$, and $\pm 1$ for the two Majorana states, a combination which is free of anomalies. In order to prohibit mixing between the two fermions, we impose a discrete symmetry: $\chi_1\to\chi_1, \, \chi_2 \to -\chi_2$. After spontaneous symmetry breaking the relevant terms of the Lagrangian are given by:
\begin{eqnarray}
	 \mathcal{L} &\supset&   \frac{1}{2} \sum_{i=1}^2 \left[ i \bar{\chi_i} \left(\slashed{\partial}  - m_{\chi_i} \right) \chi_i \pm g_D Z'^\mu\bar{\chi_i} \gamma_\mu \gamma_5 \chi_i  -\frac{m_{\chi_i}}{v_\phi} \rho \bar{\chi_i}  \chi_i \right]\, ,
\end{eqnarray}
where $m_{Z'} = 2 g_D v_\phi$.

In order to limit the number of free parameters, we assume a hierarchy in mass such that $m_{\chi_2} \gg m_{\chi_1}$, so that the abundance of $\chi_2$ is negligible compared to that of $\chi_1$. For simplicity we will rename $\chi = \chi_1$, which we identify as our dark matter candidate.

\begin{table}[t]
\begin{center}
\begin{tabular}{cccc}
\hline \hline
$m_{Z'}$ & $m_{\rho}$ & $\rho$ Decay & Main Annihilation Channels  \\
\hline \hline
 $2 m_\chi>m_{Z'}>m_\chi$ & $m_\rho < m_\chi, 2 m_{Z'} $ & $\rho\to SM$ & $\chi \chi \to Z' \rho $   \\
$m_{Z'}<m_\chi$ & $2m_{Z'}>m_\rho > m_\chi$  &$\rho\to SM$ & $\chi \chi \to Z' \rho $ and $\chi \chi \to Z' Z' $  \\
 $m_{Z'}<m_\chi$ & $m_\rho > m_\chi, 2 m_{Z'} $ &$\rho\to Z'Z' \to SM $ & $\chi \chi \to Z' \rho$ and $\chi \chi \to Z' Z' $   \\
\hline \hline
\end{tabular}
\end{center}
\caption{A summary of the distinctive regions of parameter space in a model with a hidden sector which contains dark matter in the form of a Majorana fermion, $\chi$, along with an additional vector, $Z'$, and scalar, $\rho$.}
\label{tab:MajoranaDM}
\end{table}

There are several distinctive regions of parameter space within this model, which we summarize in Table~\ref{tab:MajoranaDM}. We note that this model has been previously studied within the context of the Galactic Center Excess in Ref.~\cite{Cline:2014dwa}, although they restricted themselves to the case of $\chi \chi \to Z' Z'$. Also, the authors of Refs.~\cite{Bell:2016fqf} and~\cite{Duerr:2016tmh} have recently considered this two-portal scenario, although not within the context of the Galactic Center Excess. 

If the $Z'$ is light enough that the process $\rho \to Z'Z'$ is kinematically allowed, it will be the dominant decay channel for the dark scalar, with a width given by:
\begin{eqnarray}
 	 \Gamma(\rho \to Z' Z')  = \frac{m_\rho^3}{32
	\pi  v_\phi^2 } \left(1 -4 \frac{m_{Z'}^2}{m_\rho^2}+12
	\frac{m_{Z'}^4}{m_\rho^4}\right)  \sqrt{1-\frac{4m_{Z'}^2}{m_\rho^2}}.
\end{eqnarray}

Although we utilize the full thermally-averaged cross section in our calculations, we present only the low-velocity annihilation cross section here due to the length of these expressions in this particular model:
\begin{eqnarray}
 	 \sigma v_{\chi \chi \to Z' Z'} = \frac{g_D^4}{ 4 \pi  m_\chi \left(m_{Z'}^2-2 m_\chi^2\right)^2}  \left[m_\chi^2-m_{Z'}^2\right]^{3/2} + \mathcal{O}(v^2) \, ,
\end{eqnarray}
\begin{eqnarray}
 	 \sigma v_{\chi \chi \to Z' \rho} = \frac{g_D^4 }{64 \pi  m_\chi^4 m_{Z'}^4}\left[m_{Z'}^4 +\left(m_\rho^2-4 m_\chi^2\right)^2-2 m_{Z'}^2 \left(4 m_\chi^2+m_\rho^2\right)\right]^{3/2}+ \mathcal{O}(v^2)
\end{eqnarray}
and
\begin{eqnarray}
	 \sigma v_{\chi \chi \to \rho \rho} &=&\frac{g_D^4 m_\chi \sqrt{m_\chi^2-m_\rho^2} \left(8 m_\chi^4-8 m_\chi^2 m_\rho^2+3 m_\rho^4\right) \, v^2}{24 \pi  m_{Z'}^4 \left(m_\rho^2-4 m_\chi^2\right)^2 \left(m_\rho^2-2 m_\chi^2\right)^4} \\
	 &\times& \bigg[288 m_\chi^8-352 m_\chi^6 m_\rho^2+200 m_\chi^4 m_\rho^4-64 m_\chi^2 m_\rho^6+9 m_\rho^8\bigg]  \, . \nonumber
\end{eqnarray}

The $Z'$ induces a spin-dependent scattering cross section between our Majorana dark matter candidate and nuclei:
\begin{eqnarray}
\sigma^\text{Majorana}_\text{nucleon, SD} = \frac{3  g_D^2 g_{a}^u {}^2 \mu_{\chi N}^2 }{4 \pi m_{Z'}^4}\, ,
\end{eqnarray}
where $g_a^u$ is the axial coupling of the up quark to the $Z'$, as defined in Eq~\ref{couplingseq}. Due to the comparatively weak constraints on spin-dependent scattering and the smallness of the axial coupling in the low  $m_{Z'}$ regime, the value of $\epsilon$ is not significantly restricted in this mode.
  
Additionally, the dark matter will also experience a spin-independent interaction with nuclei as a result of Higgs exchange, with a cross section that is given by:
\begin{eqnarray}
\sigma^\text{Majorana, SI}_{\rm nucleon} =  \sin^2 2\theta \,
\frac{f_N^2}{4 \pi }  \,  \frac{\mu_{\chi N}^2 m_\chi^2 m_N^2 }{v_H^2
v_\phi^2} \left(\frac{1}{m_\rho^2} - \frac{1}{m_h^2} \right)^2 \, ,
\end{eqnarray}
leading to similar constraints on $\sin \theta$ as those shown in Fig.~\ref{fig:Vector_res}.

\begin{figure}[t]
\centering
\includegraphics[width=0.495\textwidth]{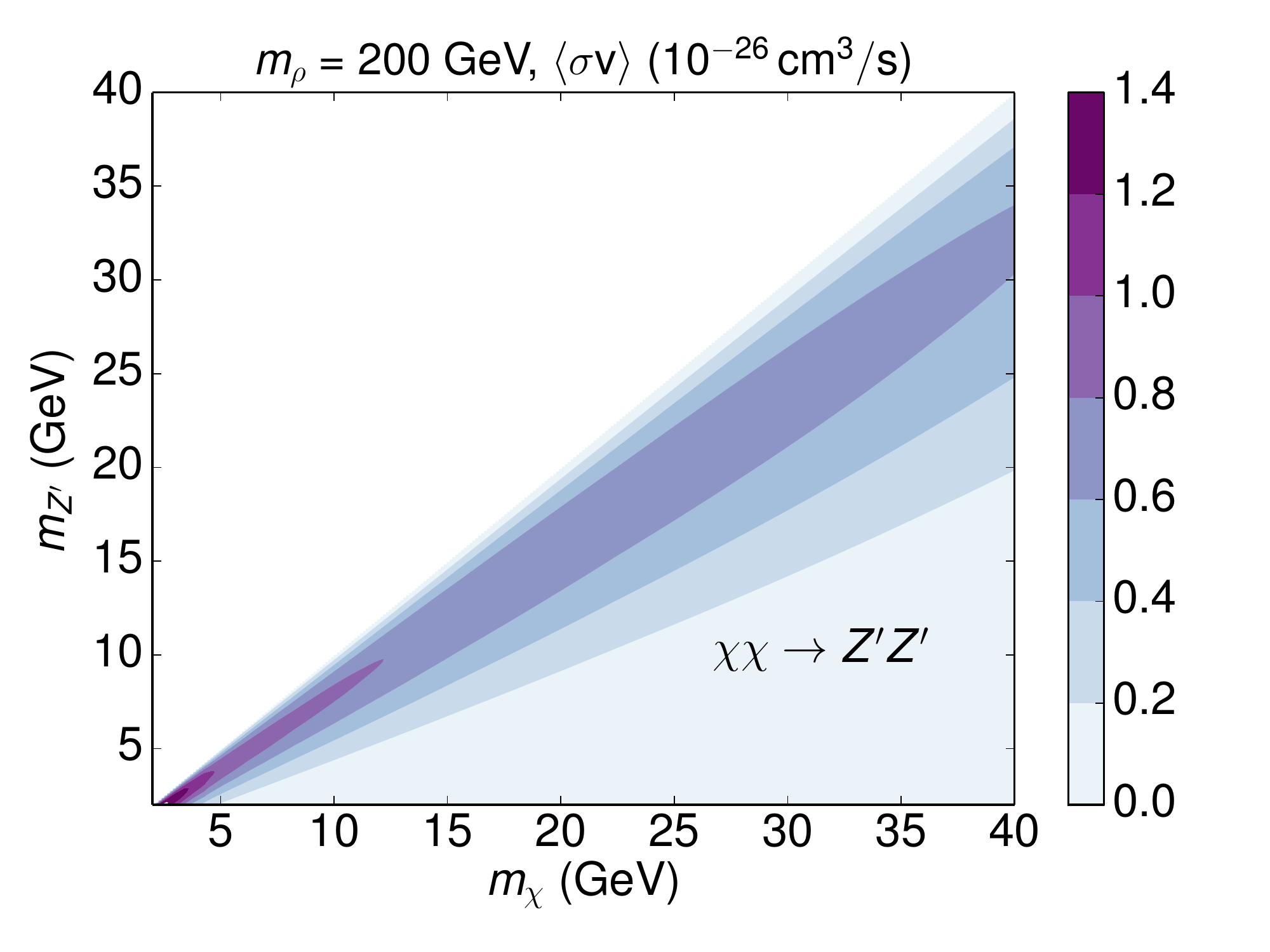}  \includegraphics[width=0.495\textwidth]{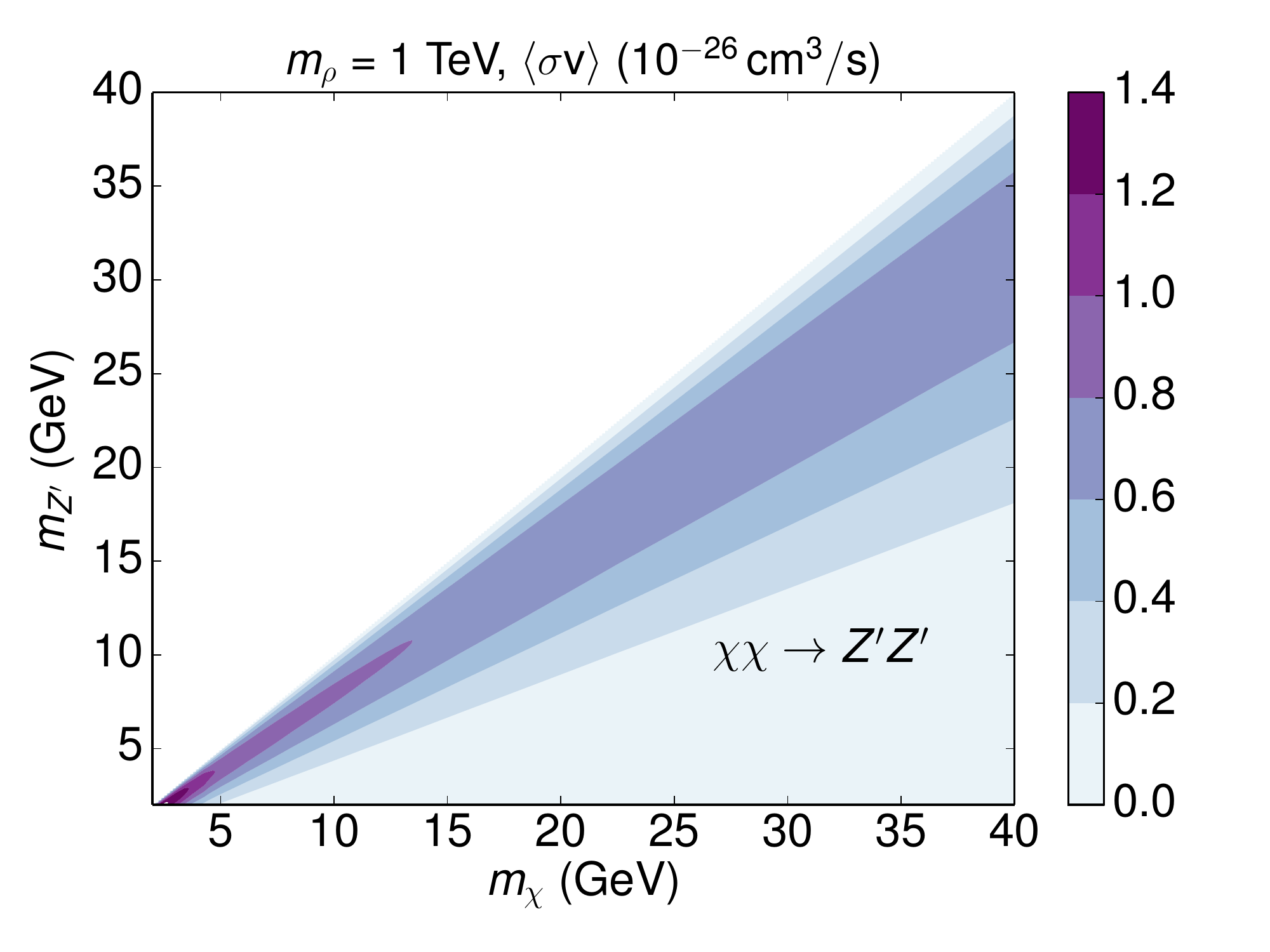}  \\
\includegraphics[width=0.495\textwidth]{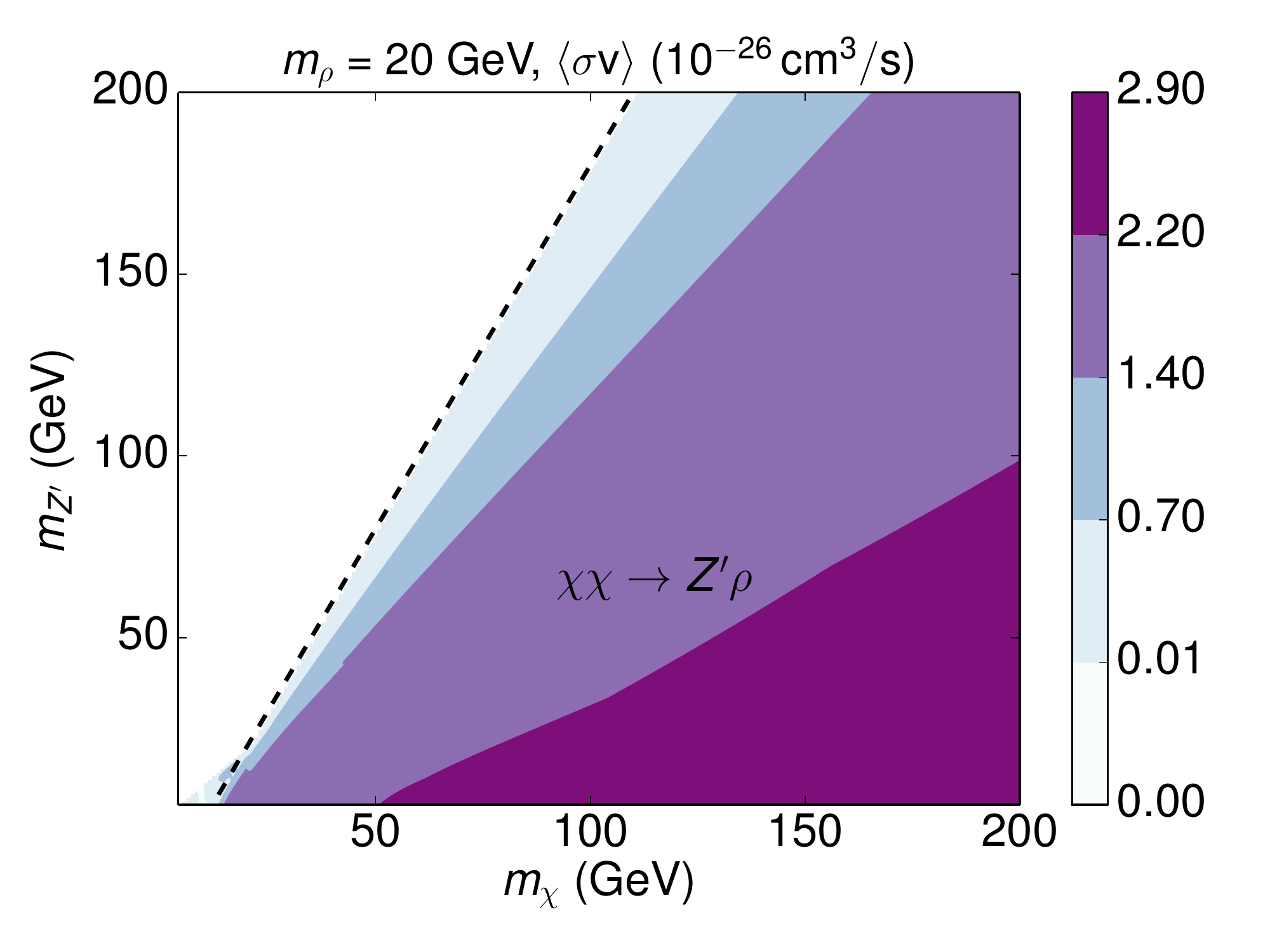} 
\includegraphics[width=0.495\textwidth]{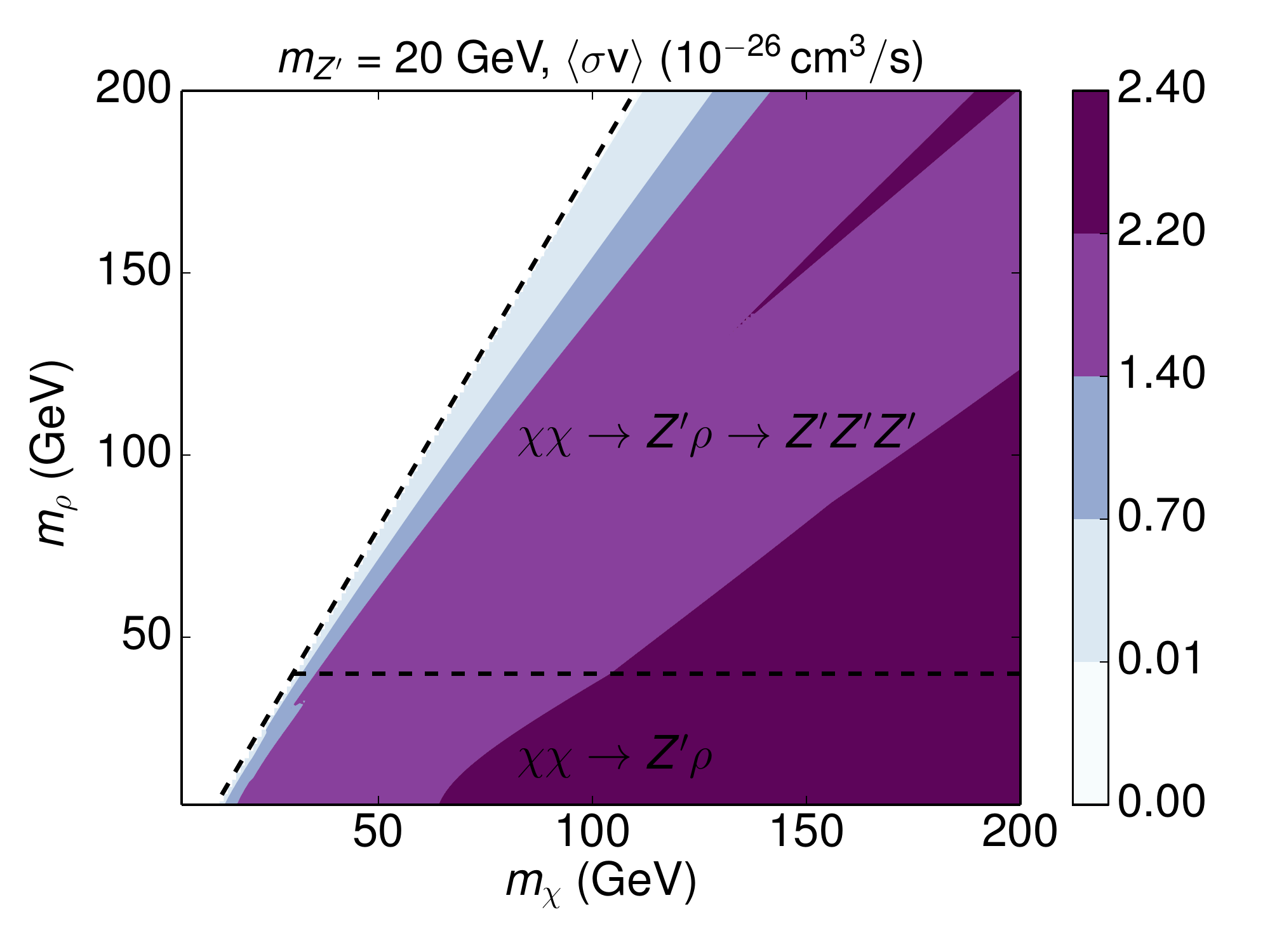}  \\
\includegraphics[width=0.495\textwidth]{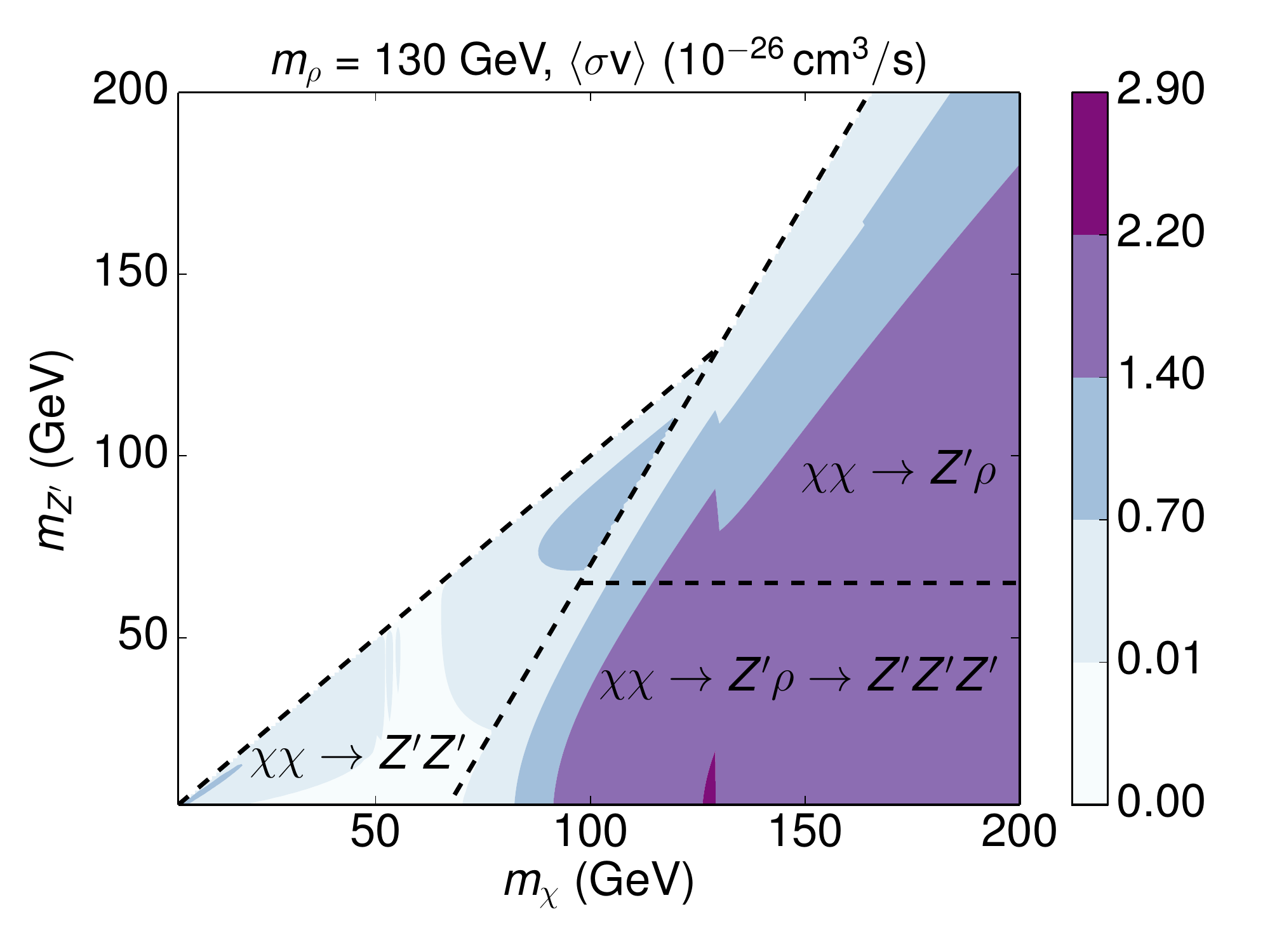} 
\includegraphics[width=0.495\textwidth]{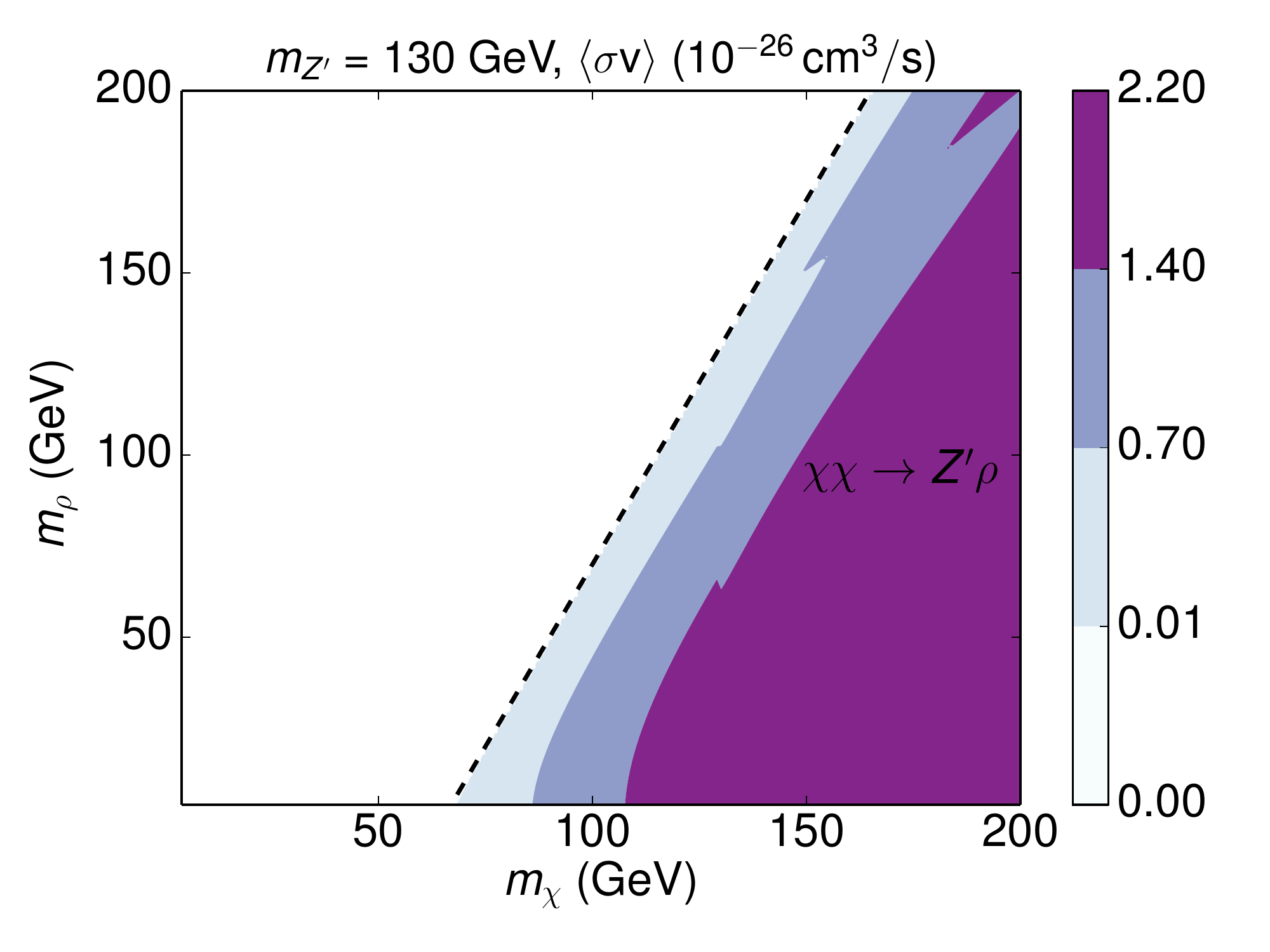}  
\caption{The annihilation cross section evaluated at a velocity of $v=10^{-3} c$ for dark matter in the form of a Majorana fermion that resides within a hidden sector which contains a vector, $Z'$, and a scalar $\rho$. In the upper frames, we show results for two cases in which $XX \rightarrow Z' Z'$ is the only kinematically allowed annihilation channel. In the middle and lower frames, a combination of the processes $XX \rightarrow Z' Z'$ and $XX \rightarrow Z' \rho$ are allowed. Whenever the $Z' \rho$ channel is open, it dominates the annihilation final state. The dashed lines demark the regions of parameter space in which different channel are kinematically allowed. Throughout each frame, the value of the hidden sector gauge coupling, $g_D$, was chosen in order to obtain a thermal relic abundance equal to the measured cosmological dark matter density.}\label{fig:MajoranaZZ_res}
\end{figure}

In Fig.~\ref{fig:MajoranaZZ_res} we plot the dark matter annihilation cross section in this model, as evaluated at a velocity of $v=10^{-3} c$, as appropriate for indirect searches, for several values of the hidden sector $\rho$ and $Z'$ masses. Throughout each frame, we have chosen the value of the coupling, $g_D$, in order to obtain a thermal relic abundance equal to the measured cosmological dark matter density. In the upper frames, we note that thermal effects which depend on the mass of the hidden sector scalar can lead the low-velocity annihilation cross section to be significantly smaller than that naively expected of a thermal relic. Once again we have assumed that either $\epsilon$ or $\sin^2 \theta$ is large enough to maintain kinetic equilibrium between the hidden sector and the Standard Model at the time of freeze-out. If this is not the case, the low-velocity annihilation cross section could be either larger or smaller than those shown in Fig.~\ref{fig:MajoranaZZ_res}~\cite{Berlin:2016vnh,Berlin:2016gtr}.

\section{Fitting the Spectrum of the Galactic Center Excess}
\label{fitting}

In this section, we calculate the gamma-ray spectrum from dark matter annihilations in the above described hidden sector models and determine the parameter space within these models that is capable of generating the observed features of the Galactic Center gamma-ray excess. 

The gamma-ray flux from dark matter annihilations is described by the following:
\begin{equation}
\Phi(E_{\gamma}, b, \ell) = \frac{\langle \sigma v \rangle}{8 \pi m_\chi^2} \frac{d\bar{N}_{\gamma}}{dE_\gamma}(E_{\gamma}) \frac{1}{\Delta \Omega} \int_{\rm los} \int_{\Delta \Omega} \, d\Omega \, ds \, \rho^2(r(s, b, \ell)) \, ,
\end{equation}
where $\langle \sigma v \rangle$ is the thermally averaged annihilation cross section, $m_\chi$ is the mass of the dark matter particle, ${d\bar{N}_{\gamma}}/{dE_\gamma}$ is the gamma-ray spectrum produced per annihilation and $\rho(r)$ is the dark matter density profile. The integrals are carried out over the observed line-of-sight (los), $s$, and over a segment of the sky of solid angle $\Delta \Omega$, denoted in terms of Galactic coordinates $b$ and $\ell$. 

Throughout this study, we will adopt a dark matter density distribution that is described by a generalized Navarro-Frenk-White (NFW) profile:
\begin{equation}
\rho(r) = \rho_s \frac{\left({r}/{r_s} \right)^{-\gamma}}{\left(1 + {r}/{r_s} \right)^{3-\gamma}} \, ,
\end{equation}
where $r$ is the distance to the Galactic Center. 
Unless otherwise stated, we adopt $\gamma = 1.2$, $r_s = 20$ kpc, and we fix $\rho_s$ by requiring that the dark matter density at a distance of 8.5 kpc from the Galactic Center is equal to $\rho_{\oplus}=0.4$ GeV/cm$^3$.

The function $d\bar{N}_{\gamma}/dE_{\gamma}$ depends on the mass of the dark matter particle and its dominant annihilation channels. For each model, the gamma-ray spectrum is given by a sum over the decay channels of the intermediate state hidden sector particle(s).
We emphasize that the gamma-ray spectrum from dark matter annihilations in hidden sector models depends not only on the dark matter mass and annihilation channels, but also on the mass of the intermediate state particles. 

Following Ref.~\cite{Elor:2015tva}, one can write the spectrum of gamma rays in the dark matter rest frame in terms of the gamma-ray spectrum in the rest frame of the intermediate state particle, $\phi$:
\begin{equation}
\label{integral}
\frac{d\bar{N}_\gamma}{dE_{\gamma}} =\frac{ 2}{m_\chi}\int_{-1}^1 \, d\cos\theta \int_{0}^1 \, dx^\prime \left(\frac{dN}{dx^\prime}\right)_\phi \delta(2x - x^\prime - \cos\theta x^\prime \sqrt{1-\kappa^2}) \, ,
\end{equation}
where $x \equiv E_{\gamma} / m_\chi$, $E_\gamma$ is the photon energy in the frame of the dark matter, $x^\prime \equiv 2 E_\gamma^\prime / m_\phi$, $E_\gamma^\prime$ is the photon energy in the frame of $\phi$, and $\kappa \equiv m_\phi / m_\chi$. The angular integration can be performed analytically, leading to
\begin{equation}\label{eq:casspec2}
\frac{d\bar{N}_\gamma}{dE_{\gamma}} = \frac{2}{m_\chi} \int_{t_{\rm min}}^{t_{\rm max}} \frac{dx^\prime}{x^\prime \sqrt{1-\kappa^2}} \left(\frac{dN}{dx^\prime} \right)_\phi\, ,
\end{equation}
where the integration limits are defined as
\begin{equation}
t_{\rm max} = \min \left[ 1 \, , \frac{2 x}{\kappa^2}\left(1+\sqrt{1-\kappa^2}\right)\right], \,\,\, t_{\rm min} = \frac{2 x}{\kappa^2}\left(1-\sqrt{1-\kappa^2}\right) \, .
\end{equation}
In the limit of small $\kappa$, Eq.~\ref{eq:casspec2} reduces to

\begin{equation}\label{eq:casspec3}
\frac{d\bar{N}_\gamma}{dE_{\gamma}} = \frac{2}{m_\chi} \int_{x}^{1} \frac{dx^\prime}{x^\prime}\left(\frac{dN}{dx^\prime} \right)_\phi \, .
\end{equation}

A publicly available code for the calculation of cascade spectra, using the direct annihilation spectrum tabulated in Ref.~\cite{Cirelli:2010xx}, has been presented and described in Ref.~\cite{Elor:2015tva,Elor:2015bho}. The spectrum used for the analysis of $\chi \chi \rightarrow Z' Z'$ (with $m_{Z'} > 2\, \,{\rm GeV}$) and $\chi \chi \rightarrow \rho \rho$ were produced using this code. This code does not, however, include annihilations to mesons, as is required in the case of a light $Z'$, nor does it allow for the immediate computation of the spectrum that arises from annihilations featuring asymmetric boosts, such as in the case of $\chi \chi \rightarrow Z' \rho$.  In the case of $m_{Z'} \lsim 2$ GeV, the calculation of the branching fractions is non-trivial due to the appearance of hadronic resonances which invalidate the QCD description in terms of final state quarks. At masses $m_{Z^\prime} \lsim 1.5$ GeV, the $Z'$ decays predominately to electrons, muons, and a small number of hadronic resonances. In this low-mass regime, we adopt the branching fractions as presented in Ref.~\cite{Buschmann:2015awa} and utilized PYTHIA~8~\cite{Sjostrand:2007gs} to obtain the gamma-ray spectra from the decays of the $Z'$ to mesons. The cascade spectra for these decay channels was then computed using Eq.~\ref{eq:casspec3}. The code of Refs.~\cite{Elor:2015tva,Elor:2015bho} was used to compute the $e^+e^-$ and $\mu^+\mu^-$ spectrum for this decay.  For the case of annihilations to multiple intermediate state particle species (\ie $\chi \chi \rightarrow Z' \rho$) we generalized Eq.~\ref{integral} to asymmetric annihilations using the results of Ref.~\cite{Berlin:2014pya} and utilized this result.

We note that although the above expressions were derived in Ref.~\cite{Elor:2015tva} for the case of scalars, it has been shown that the spectrum arising from the decay of vectors with reasonable angular dependencies leads to similar modifications to the spectrum. Thus we treat the vector portal models using the same formalism as those of the Higgs portal.

To assess whether a given dark matter model is capable of generating the observed features of the Galactic Center gamma-ray excess, we utilize the results of the Fermi data analysis carried out by Calore, Cholis and Weniger~\cite{Calore:2014xka}. More specifically, we extract the data points, statistical errors, and the first three principal components of the decomposition of the covariance matrix of residuals, as shown in Figs. 14 and 12 of Ref.~\cite{Calore:2014xka}, respectively. We then calculate the value of the $\chi^2$, which is given as follows: 
\begin{equation}
\chi^2 = \sum_{i,j} \left(\frac{d\bar{N}_{\gamma}}{dE_{\gamma,i}}({\bf \Theta}) - \frac{dN_{\gamma}}{dE_{\gamma,i}}\right)\Sigma_{ij}^{-1}\left(\frac{d\bar{N}_{\gamma}}{dE_{\gamma, j}}({\bf \Theta}) - \frac{dN_{\gamma}}{dE_{\gamma, j}}\right) \, ,
\end{equation}
where $\Sigma_{ij}$ is full covariance matrix, given by
\begin{equation}
\Sigma_{ij} = (\sigma_{i}^2)\delta_{i,j} + \Sigma_{ij, {\rm mod}}^{\rm trunc} + \Sigma_{ij, {\rm res}} \, .
\end{equation}
Here, $dN_{\gamma}/dE_{\gamma, i}$ and $d\bar{N}_{\gamma}/dE_{\gamma, i}({\bf \Theta})$ are the measured and predicted flux in bin $i$, and ${\bf \Theta}$ denotes the parameters of the dark matter model under consideration. $\Sigma_{ij, {\rm mod}}^{\rm trunc}$ is the truncated covariance matrix which accounts for empirical model systemics, approximated here (and in Ref.~\cite{Calore:2014xka}) using the first three principal components, and $\Sigma_{ij, {\rm res}}$ accounts for the residual systematics below $1$ GeV, modeled as:
\begin{equation}
\Sigma_{ij, {\rm res}} = \frac{dN}{dE_i^{\text res}}\frac{dN}{dE_j^{\text res}} + \delta_{i,j} \frac{dN}{dE_i^{\text res}}\frac{dN}{dE_j^{\text res}} \, .
\end{equation}

\begin{figure*}
\center
\includegraphics[width=.6\textwidth]{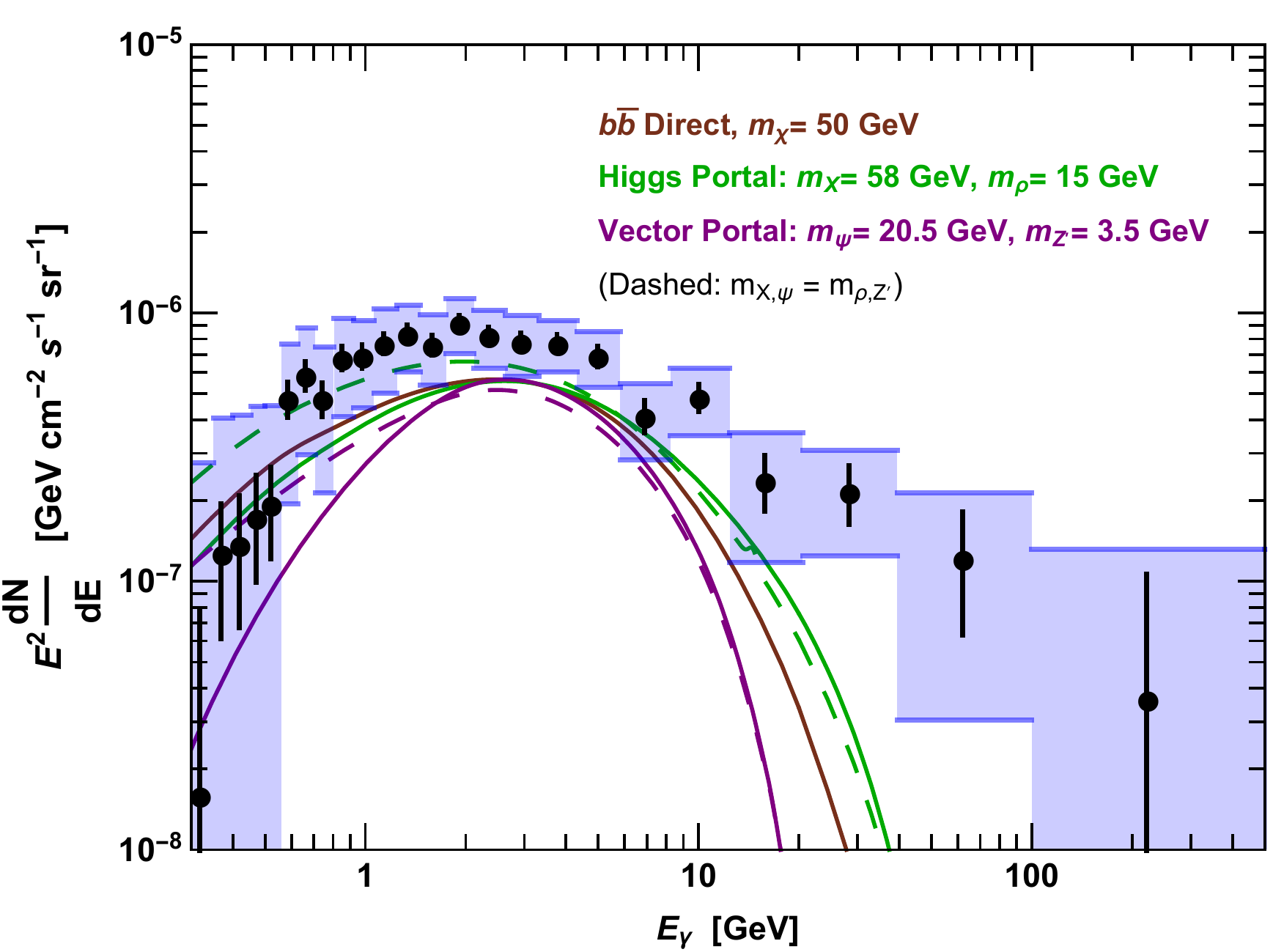} \\
\includegraphics[width=.6\textwidth]{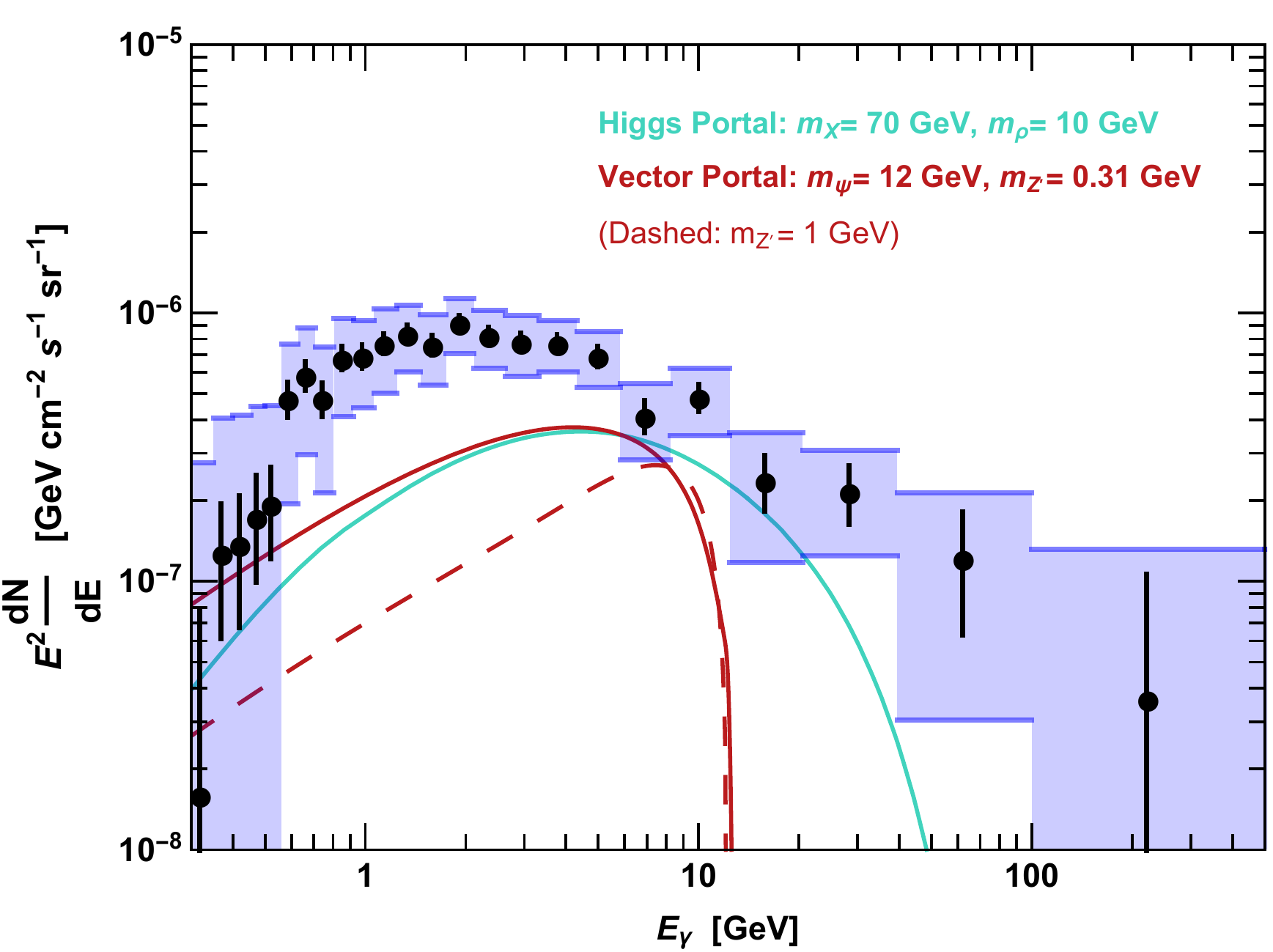}
\caption{\label{fig:gc_excess_comparison} The spectrum of the Galactic Center gamma-ray excess  as presented in Ref.~\cite{Calore:2014xka} compared with that predicted in selected dark matter models. The black error bars represent the statistical uncertainty while blue bands represent the diagonal contributions to the covariance matrix from the statistical errors and modeling systematics. Each of the models shown in the upper frame provides a good fit to the data (featuring $p$-values in the range of 0.37 to 0.46), while those in the lower frame do not (featuring $p$-values of 0.02 or less).}
\end{figure*}

In \Fig{fig:gc_excess_comparison} we compare the observed spectrum of the Galactic Center gamma-ray excess with that predicted for a selection of annihilating dark matter models. The vertical black error bars correspond to the statistical error in each bin, while the blue bars represent the diagonal contributions of the statistical errors and modeling systematics to the covariance matrix.\footnote{We exclude here the residual systemics below $1$ GeV so that a direct comparison can be made with Fig.~14 of Ref.~\cite{Calore:2014xka}.}  The upper frame of this figure illustrates that dark matter can provide a good fit to the observed spectrum if it annihilates directly to Standard Model particles (such as in the case of a 50 GeV dark matter particle annihilating directly to $b\bar{b}$, featuring a $p$-value of 0.43) and also if it instead annihilates to intermediate unstable states (such as a 58 GeV dark matter particle which annihilates to a pair of 15 or 58 GeV particles that decay through the Higgs portal, or a 20.5 GeV dark matter particle which annihilates to a pair of 3.5 or 20.5 GeV particles that decay through the vector portal, which each provide $p$-values in the range of 0.37 to 0.46).\footnote{Due to correlations between error bars, the models shown in Fig.~\ref{fig:gc_excess_comparison} generally provide a better fit to the data than may appear.} In the lower frame, we show some examples of dark matter models which do not provide a good fit to the observed gamma-ray spectrum. In particular, we find no parameter space with $m_{Z'} < 2$ GeV which provides a good fit to the data. For each dark matter model shown in Fig.~\ref{fig:gc_excess_comparison}, the value of the annihilation cross section was selected to provide the best-fit to the measured spectrum of the Galactic Center excess (adopting $\gamma = 1.2$, $R_\oplus = 8.5$ kpc, and $\rho_\oplus = 0.4$ GeV$/{\rm cm^3}$). 
\begin{figure*}
\center
\includegraphics[width=.47\textwidth]{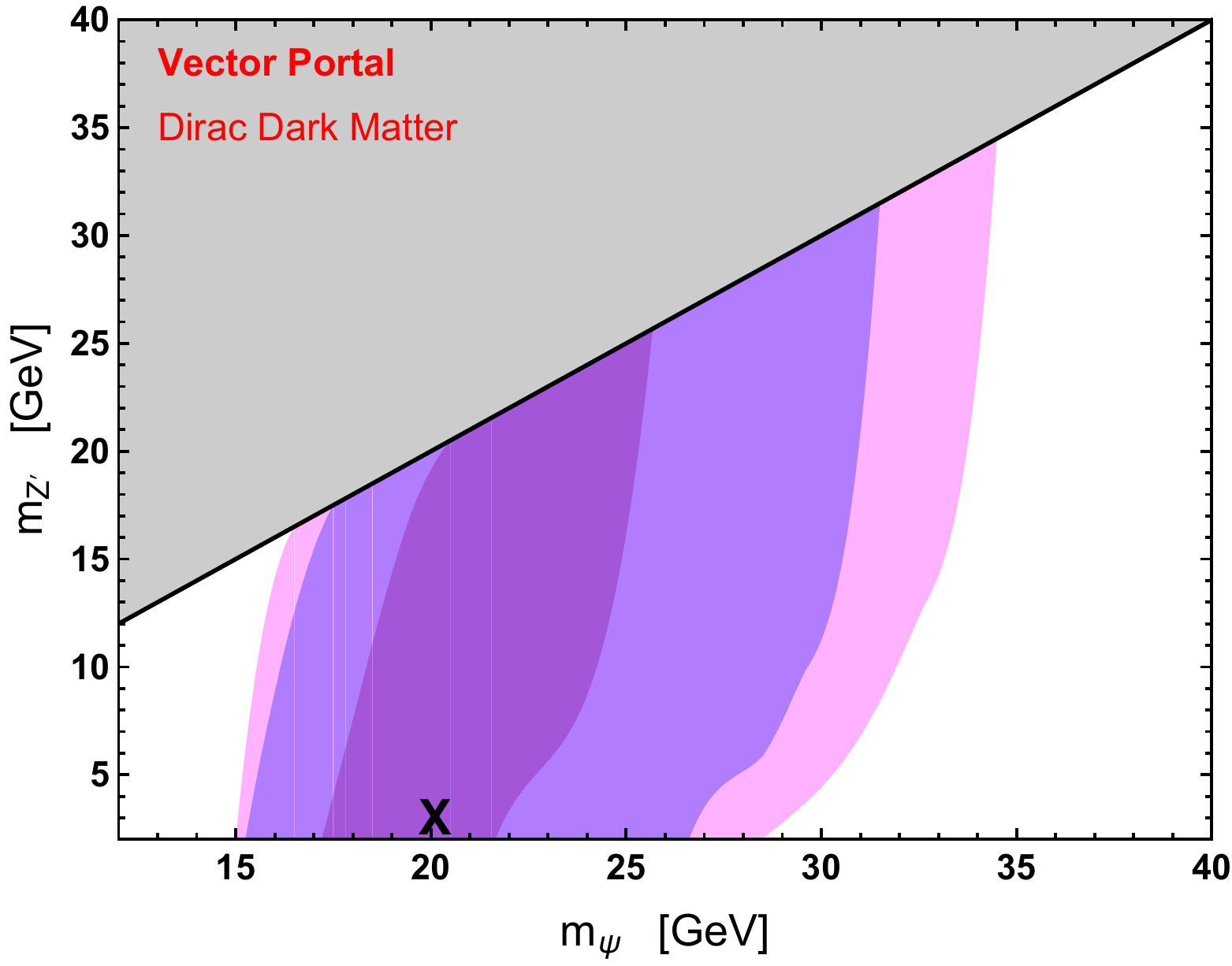}
\includegraphics[width=.47\textwidth]{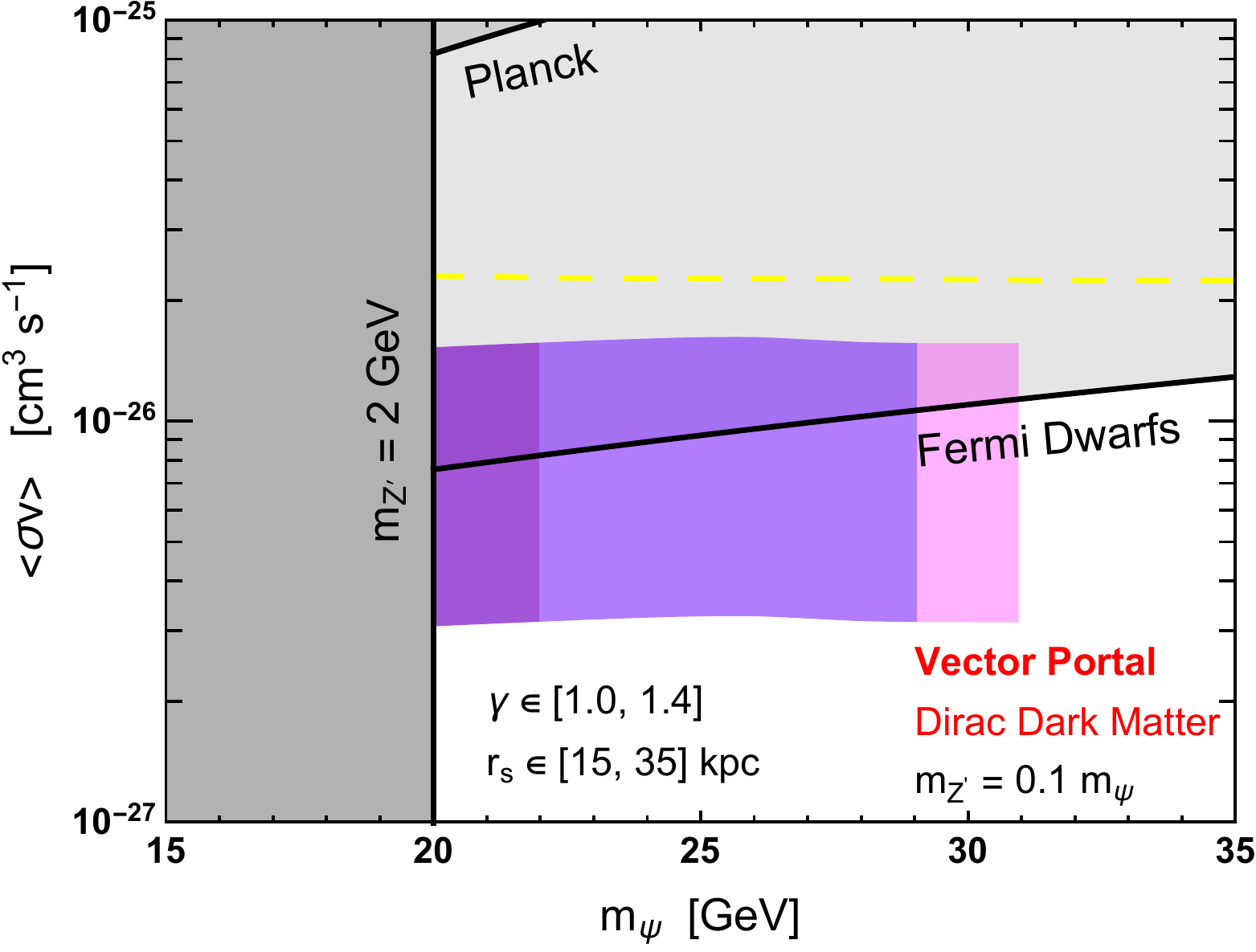} \\
\includegraphics[width=.49\textwidth]{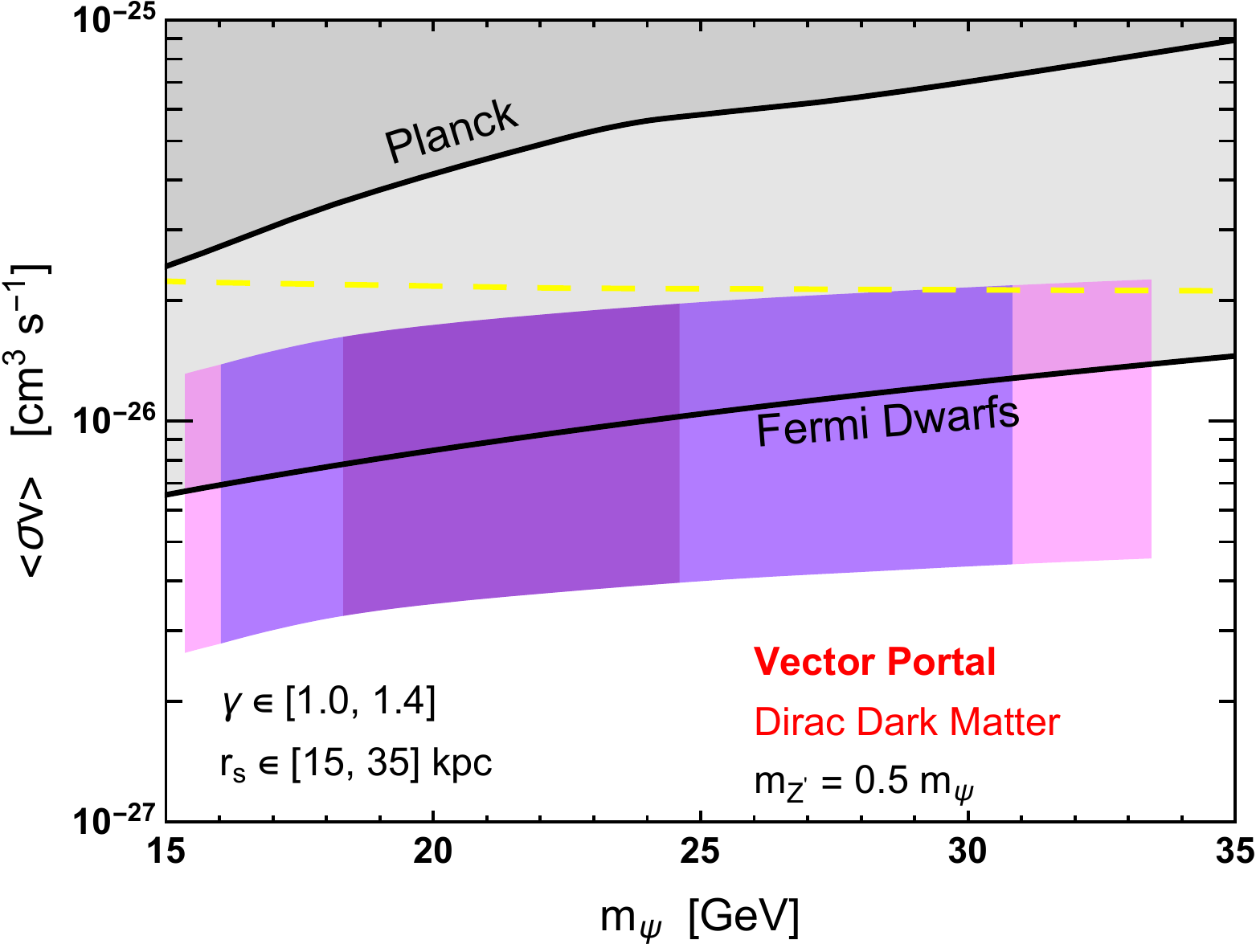}
\includegraphics[width=.49\textwidth]{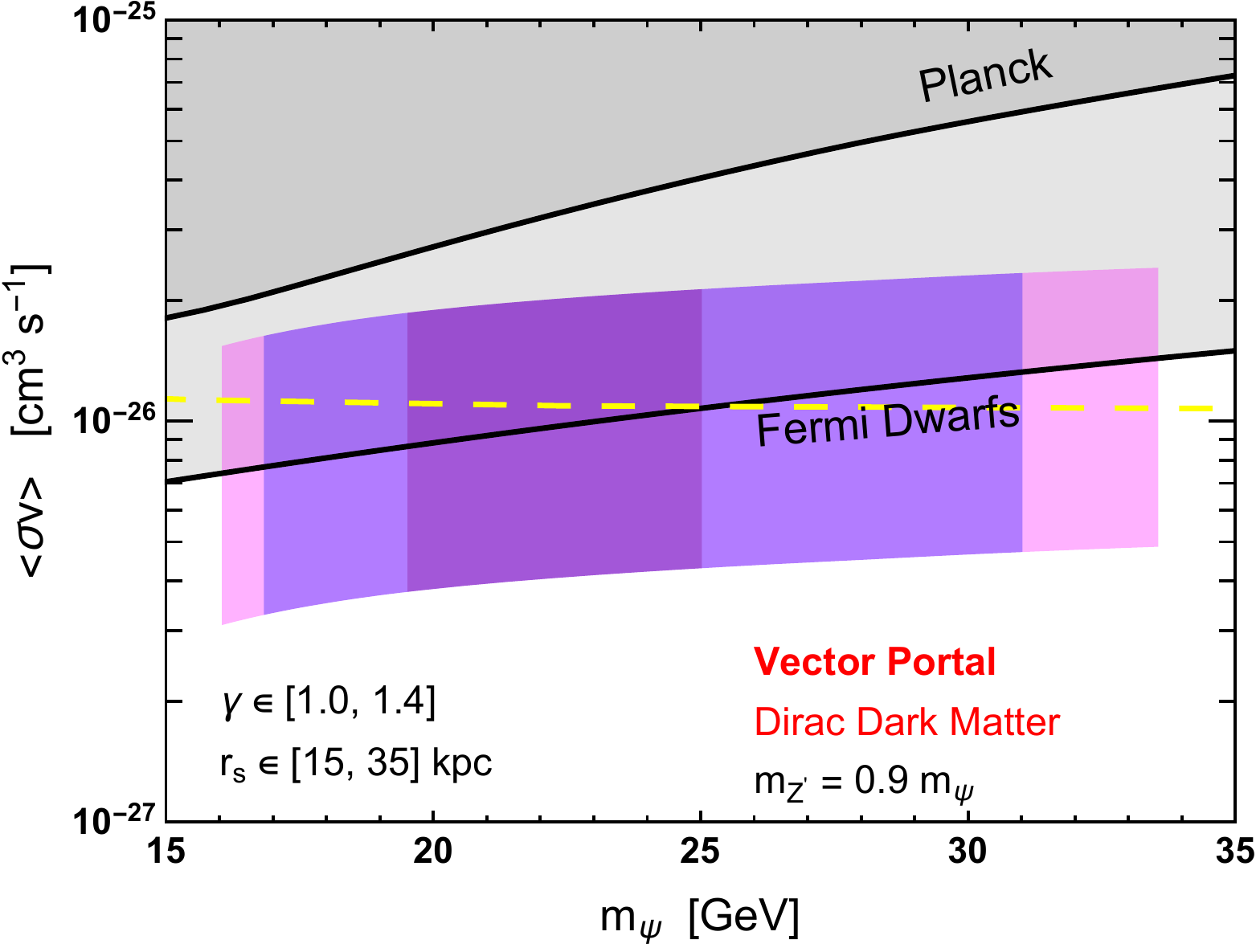}
\caption{ \label{fig:photon_param_space} Parameter space in a model with dark matter in the form of a Dirac fermion, $\Psi$, that annihilates to pairs of hidden sector vectors, $Z'$, which decay to the Standard Model through the vector portal (see Sec.~\ref{subsec:Dirac}). In the upper left frame, the bands denote the regions in which the model provides a fit to the measured Galactic Center gamma-ray excess which yields a $p-$value of $\geq 0.32$ (dark purple), $\geq 0.10$ (violet), and $\geq 0.05$ (magenta). The best-fit point is shown with an `X'. In the other three frames, we plot the best-fit range for the dark matter annihilation cross section in this model, for the case of $m_{Z'}=0.1 \, m_{\Psi}$, $0.5 \, m_{\Psi}$ or $0.9 \, m_{\Psi}$. The vertical width of the band corresponds to varying the parameters of the halo profile as indicated. Also shown are bounds from Fermi's observations of dwarf galaxies~\cite{Fermi-LAT:2016uux} and from Planck~\cite{Slatyer:2015jla,Slatyer:2015kla}. The yellow dashed lines are the contours over which the thermal relic abundance is equal to the measured cosmological dark matter density. This model can account for the entirety of the dark matter and generate the observed characteristics of the Galactic Center gamma-ray excess for $m_{\Psi} \sim 15-35$ GeV and $m_{Z'} \sim (0.8-1) \, m_{\Psi}$.}
\end{figure*}

\begin{figure*}
\center
\includegraphics[width=.48\textwidth]{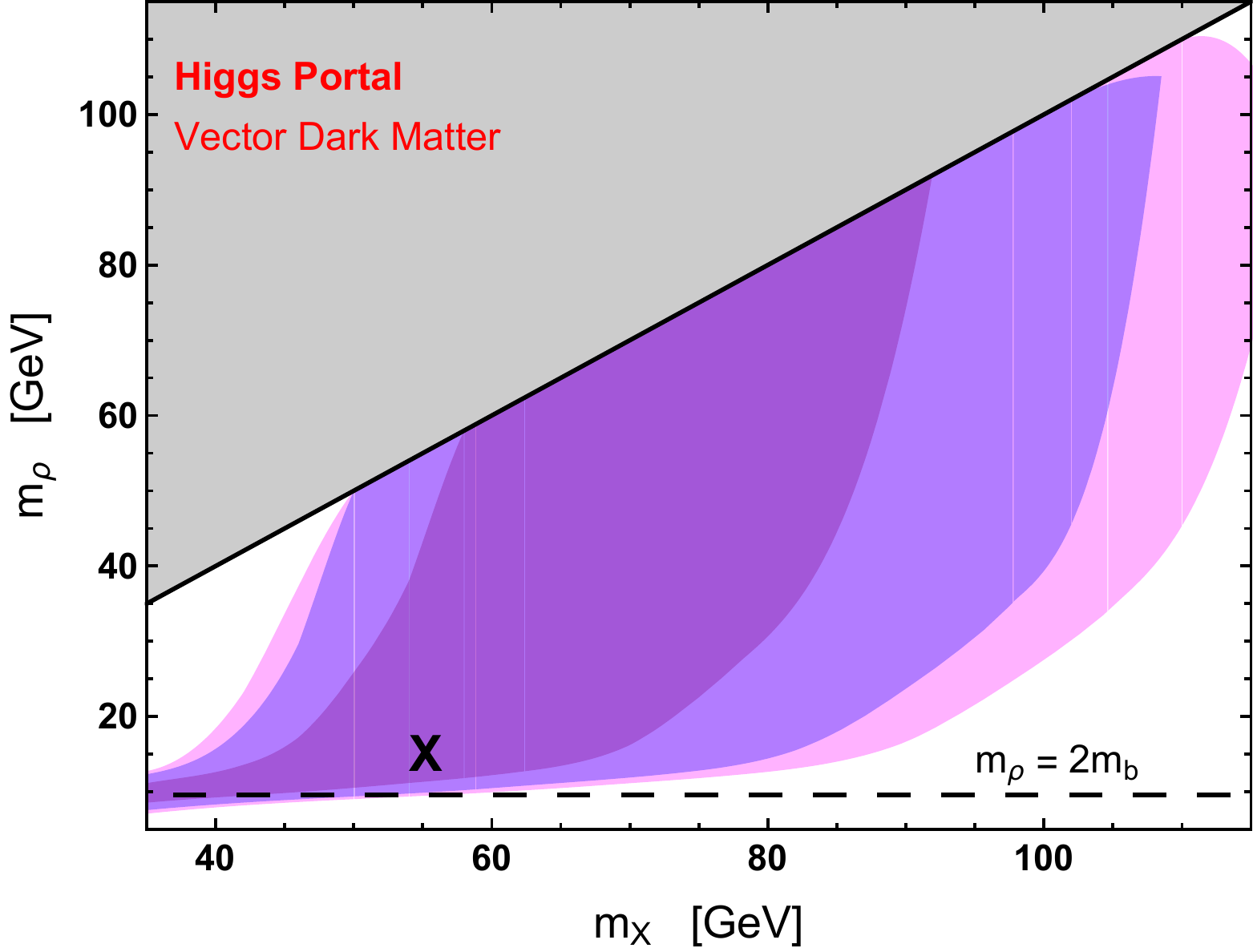} \\
\includegraphics[width=.49\textwidth]{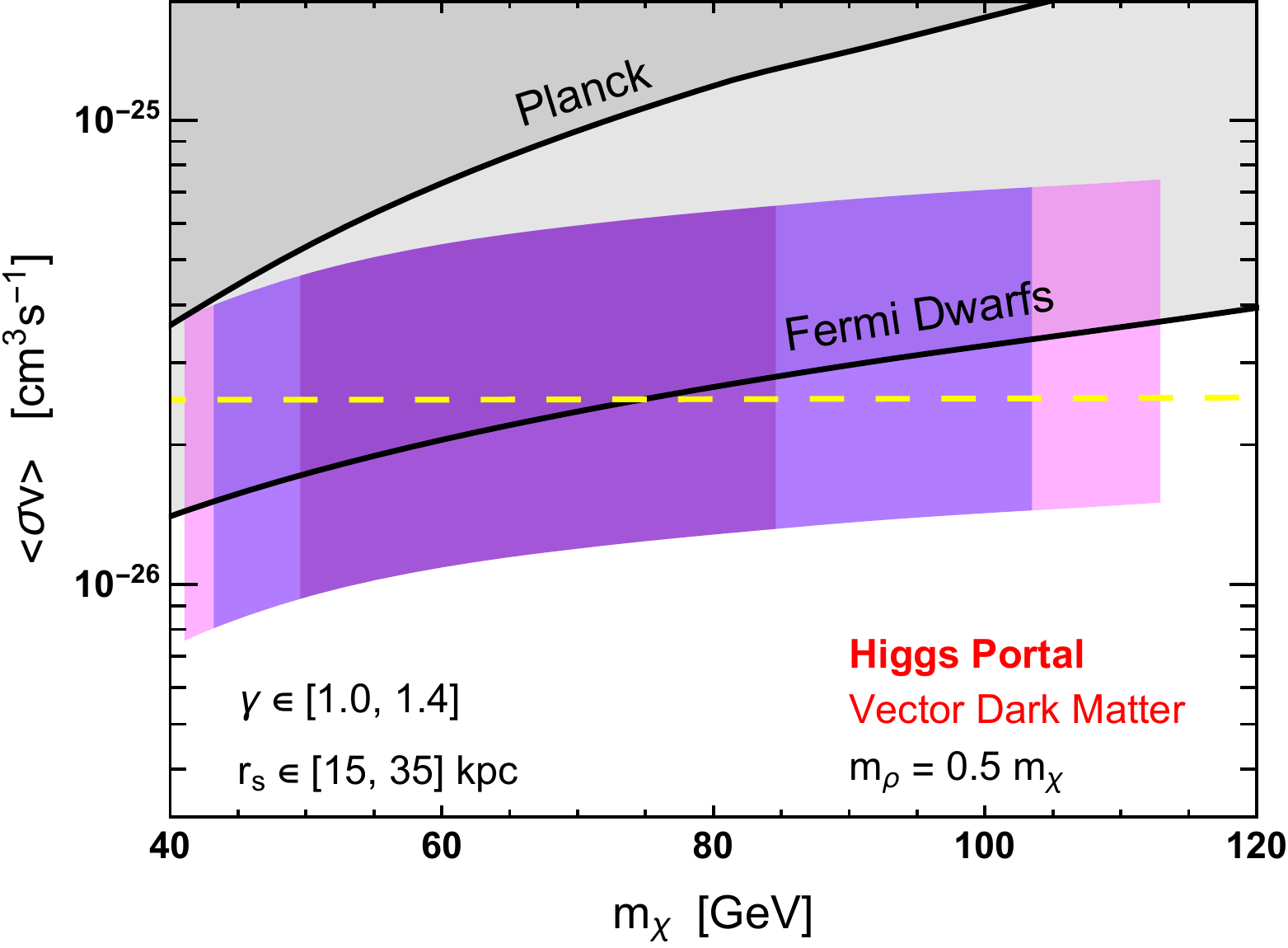}
\includegraphics[width=.49\textwidth]{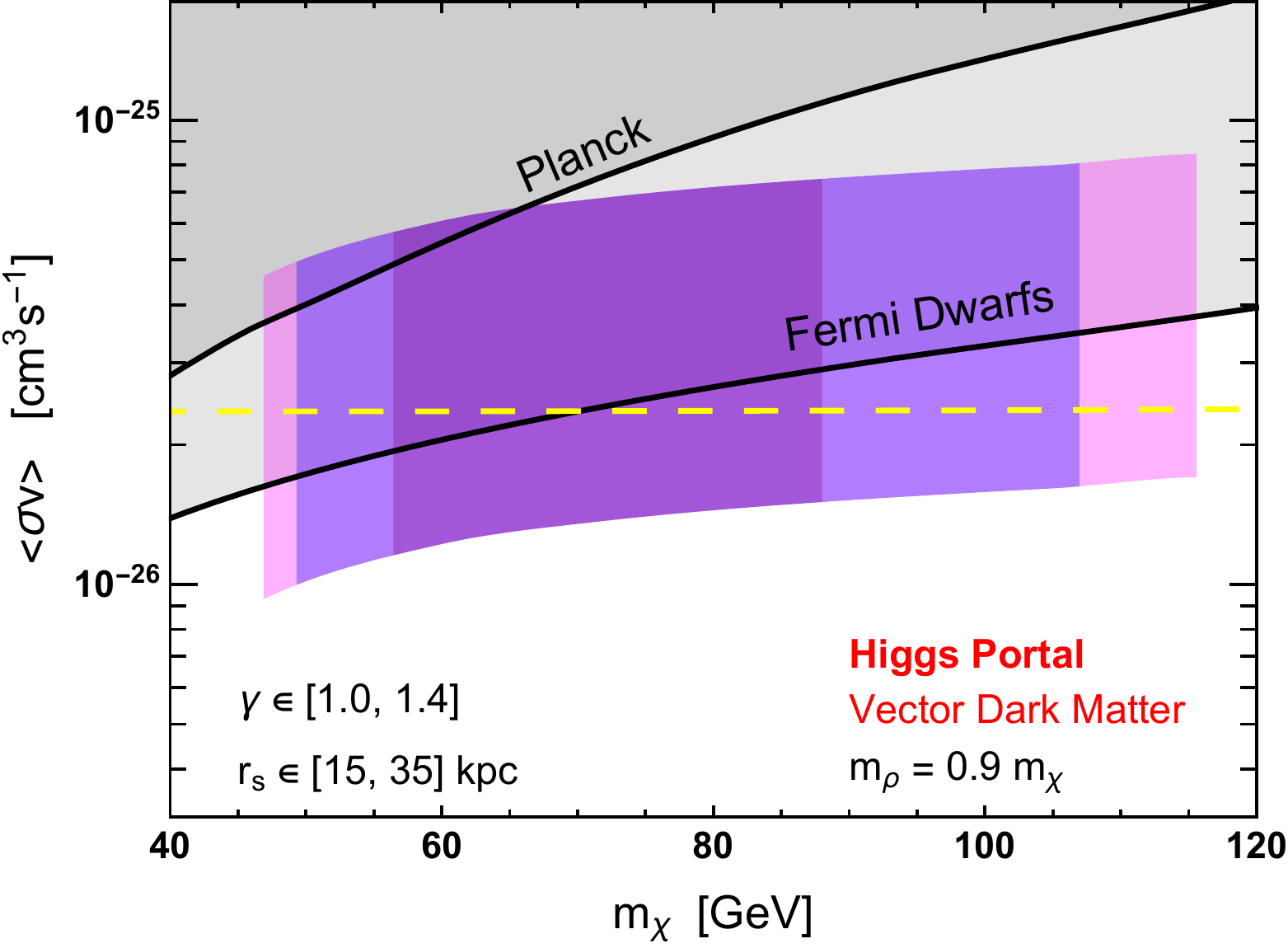}
\caption{\label{fig:higgs_param_space} As in \Fig{fig:photon_param_space}, but for dark matter in the form of a vector that annihilates to pairs of scalars which decay to the Standard Model through the Higgs portal (see Sec.~\ref{subsec:Vector}). This model can account for the entirety of the dark matter and generate the observed characteristics of the Galactic Center gamma-ray excess for $m_{X} \sim 70-110$ GeV and for a wide range of $m_{\rho}$.}
\end{figure*}

In Fig.~\ref{fig:photon_param_space}, we show results for a model with a Dirac fermion dark matter candidate, $X$, that annihilates to pairs of hidden sector vectors, $Z'$, which then decay to the Standard Model through the vector portal (see Sec.~\ref{subsec:Dirac}). We show both the parameter space that yields a good fit to the measured spectrum of the excess and the range of low-velocity annihilation cross sections that are favored by this fit. More specifically, we calculate the $\chi^2$ for the best-fit value of the annihilation cross section for each point of the parameter space. In the upper left panel, we highlight the regions which yield $p-$values $\geq 0.32$ (dark purple), $\geq 0.10$ (violet), and $\geq 0.05$ (magenta), and mark the best-fit point with an `X'. In the remaining three panels, we show the best-fit annihilation cross section, $\langle \sigma v\rangle$, as a function of the dark matter mass, for three values of $m_{Z'}/m_{\Psi}$. The vertical width of this band denotes the impact of varying $\gamma$ between $1$ and $1.4$ and $r_s$ between 15 and 35 kpc. Note that by varying the local dark matter density between $0.2$ and $0.6$ GeV/${\rm cm}^3$ this band would be further extended by a factor of 2.25 (4) downward (upward). The colors of these band have the same meaning as in the upper left panel.\footnote{The $p-$values corresponding to the bands in the right panel of \Fig{fig:higgs_param_space} only correspond to spectral fit (\ie they do not include morphological information), and thus changes in the density profile can be absorbed into the cross section without altering the $\chi^2$ value. We have chosen here to simply scan over a reasonable range of astrophysical parameters that are approximately consistent with the morphology of the observed gamma-ray excess.}  

Also shown in Fig.~\ref{fig:photon_param_space} are constraints on the dark matter annihilation cross section as derived from Fermi's observations of dwarf spheroidal galaxies~\cite{Fermi-LAT:2016uux} and from Planck.  Regarding the constraint from Planck, we follow the approach of Refs.~\cite{Slatyer:2015jla,Slatyer:2015kla} (adopting the `3 keV' prescription) to calculate the impact of dark matter annihilation on the history of recombination. The bounds from dwarf spheroidal galaxies are calculated following the statistical procedure outlined in \cite{Ackermann:2015zua} using the 19 dwarf galaxies with measured J-factors listed in Table 1 of \cite{Fermi-LAT:2016uux}. Specifically, for each model and for each choice of dark matter and mediator mass, the spectrum is calculated from $500$ MeV to $500$ GeV and compared with the precomputed bin-by-bin likelihood analyses for each dwarf provided in~\cite{dwarfOnline}. Uncertainties in the J-factor are treated with a Gaussian likelihood term as shown in Eq.~3 of \cite{Ackermann:2015zua}. In the case of constraints from dwarf galaxies, one should bear in mind that these constraints are subject to non-negligible uncertainties, such as those associated with departures from spherical symmetry~\cite{Klop:2016lug,Ichikawa:2016nbi,Sanders:2016eie} and with issues associated with stellar membership~\cite{Evans:2016xwx}.

The dashed yellow lines shown in Fig.~\ref{fig:photon_param_space} represent the parameter space which yields a thermal relic abundance equal to the measured cosmological dark matter density. In this model, we can account for the entirety of the dark matter and generate the observed characteristics of the Galactic Center gamma-ray excess for $m_{\Psi} \sim 15-35$ GeV and $m_{Z'} \sim (0.8-1) \, m_{\Psi}$, agreeing well with previous studies~\cite{Cirelli:2016rnw}.

We note that it is possible to construct hidden sector dark matter models which feature a $Z'$ with very different decay modes than those considered here. For example, motivated by recent anomalies associated with semi-leptonic $b$-decays, models have been discussed in which a $Z'$ obtains flavor non-universal couples via mixing with an additional vector-like family, leading to large couplings to third generation fermions. Although we do not explicitly study models of this variety here, we note that they can also provide a good fit to the observed characteristics of the Galactic Center gamma-ray excess~\cite{Cline:2017lvv}.

\begin{figure*}[t]
\center
\includegraphics[width=.49\textwidth]{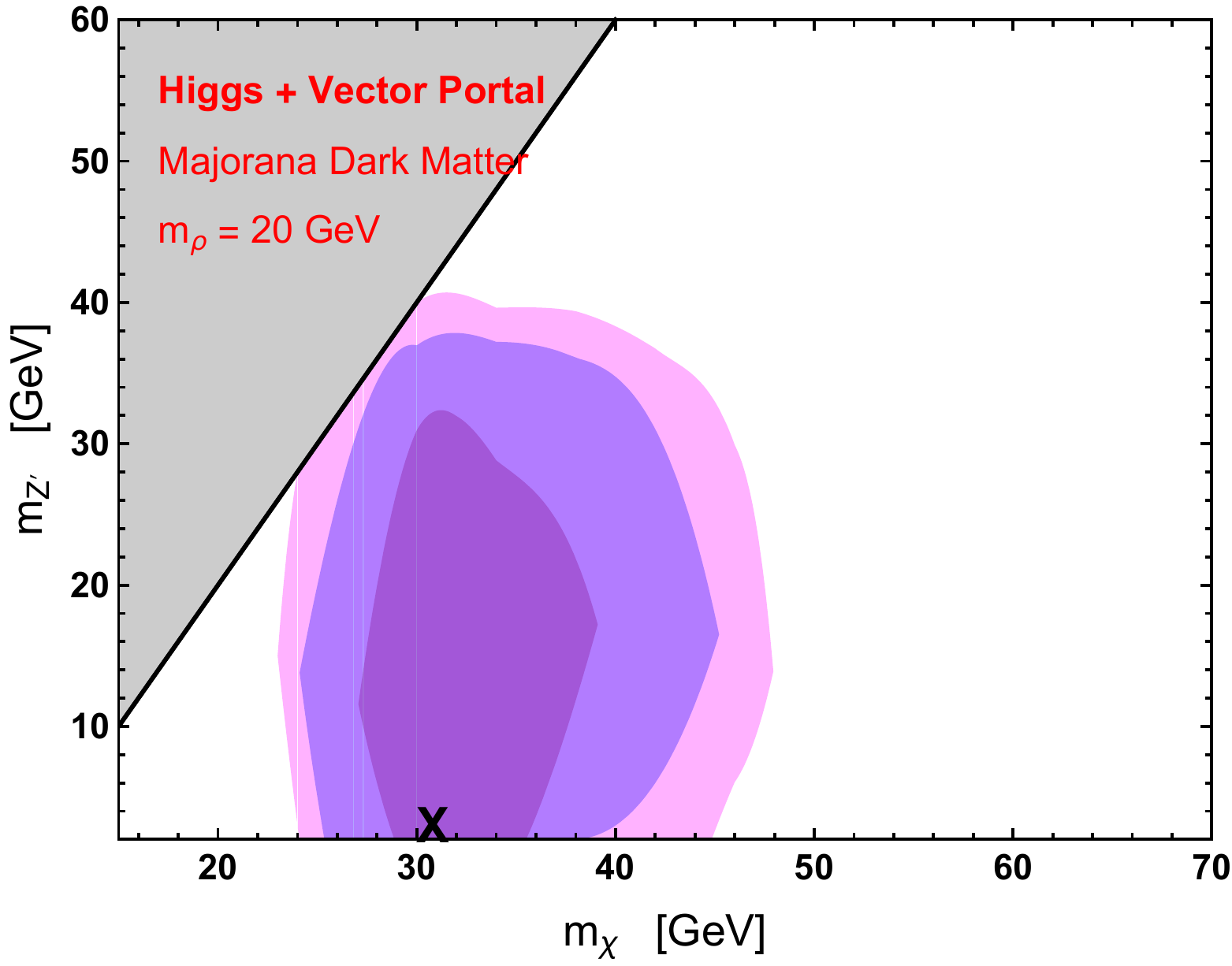}
\includegraphics[width=.49\textwidth]{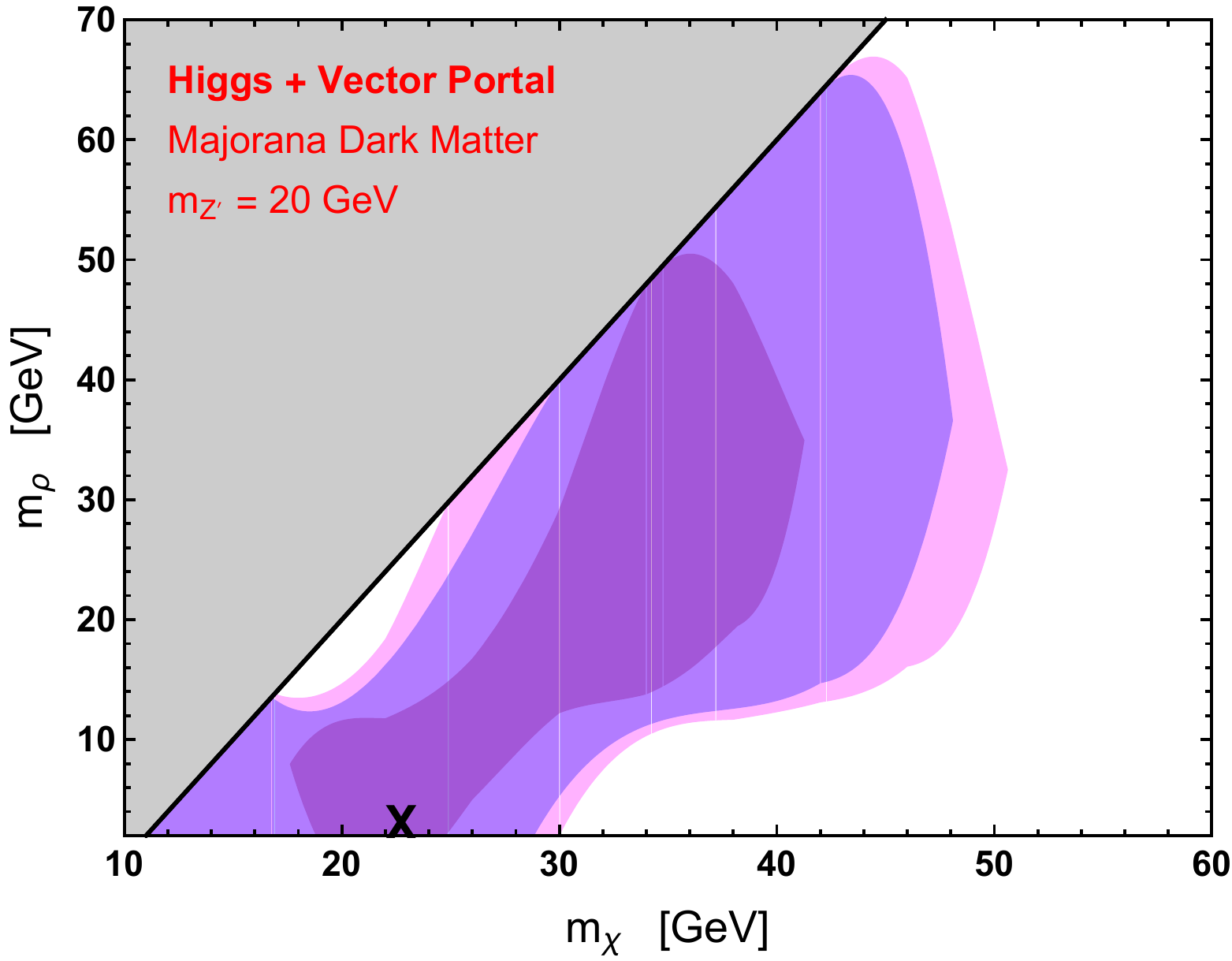}
\includegraphics[width=.49\textwidth]{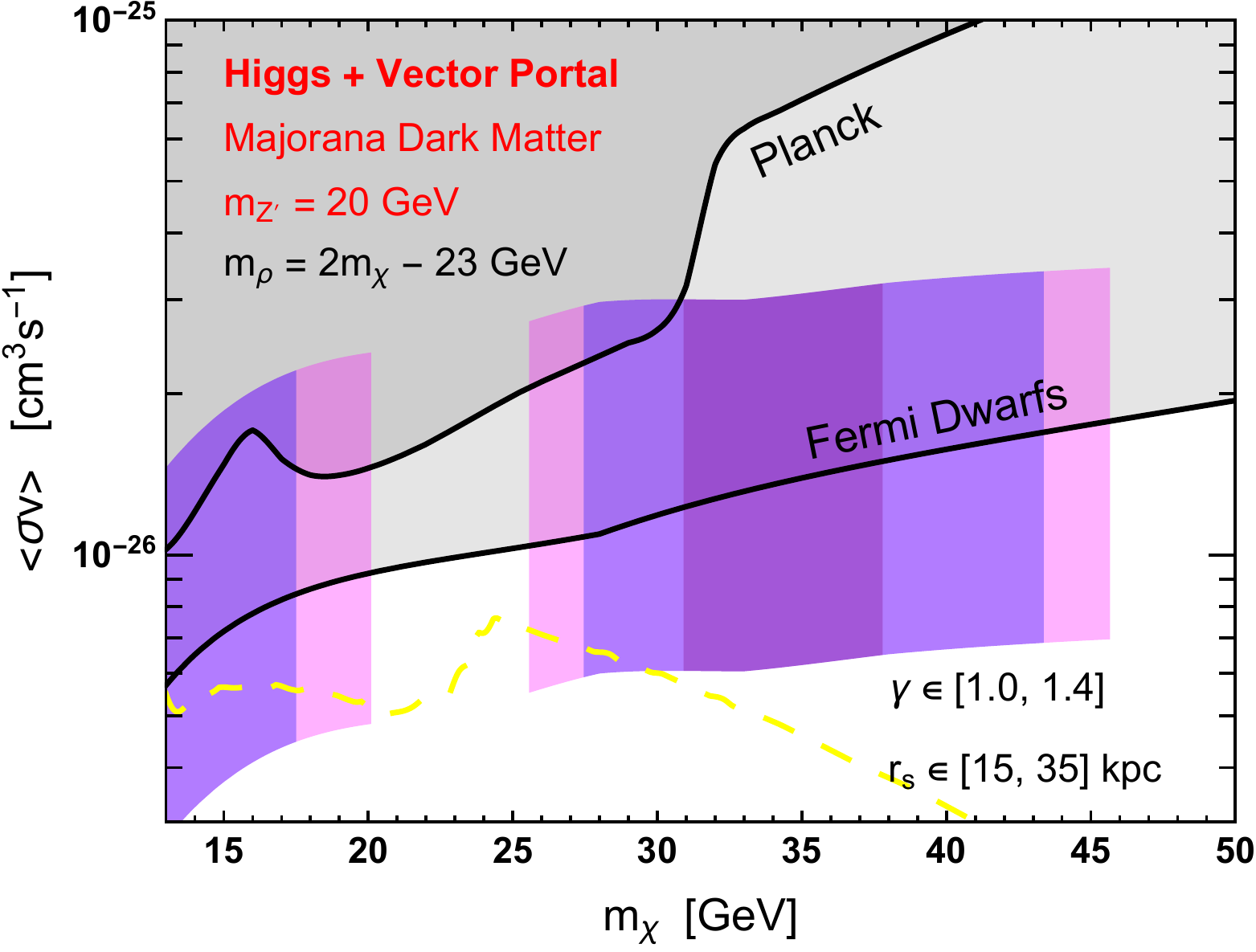}
\includegraphics[width=.49\textwidth]{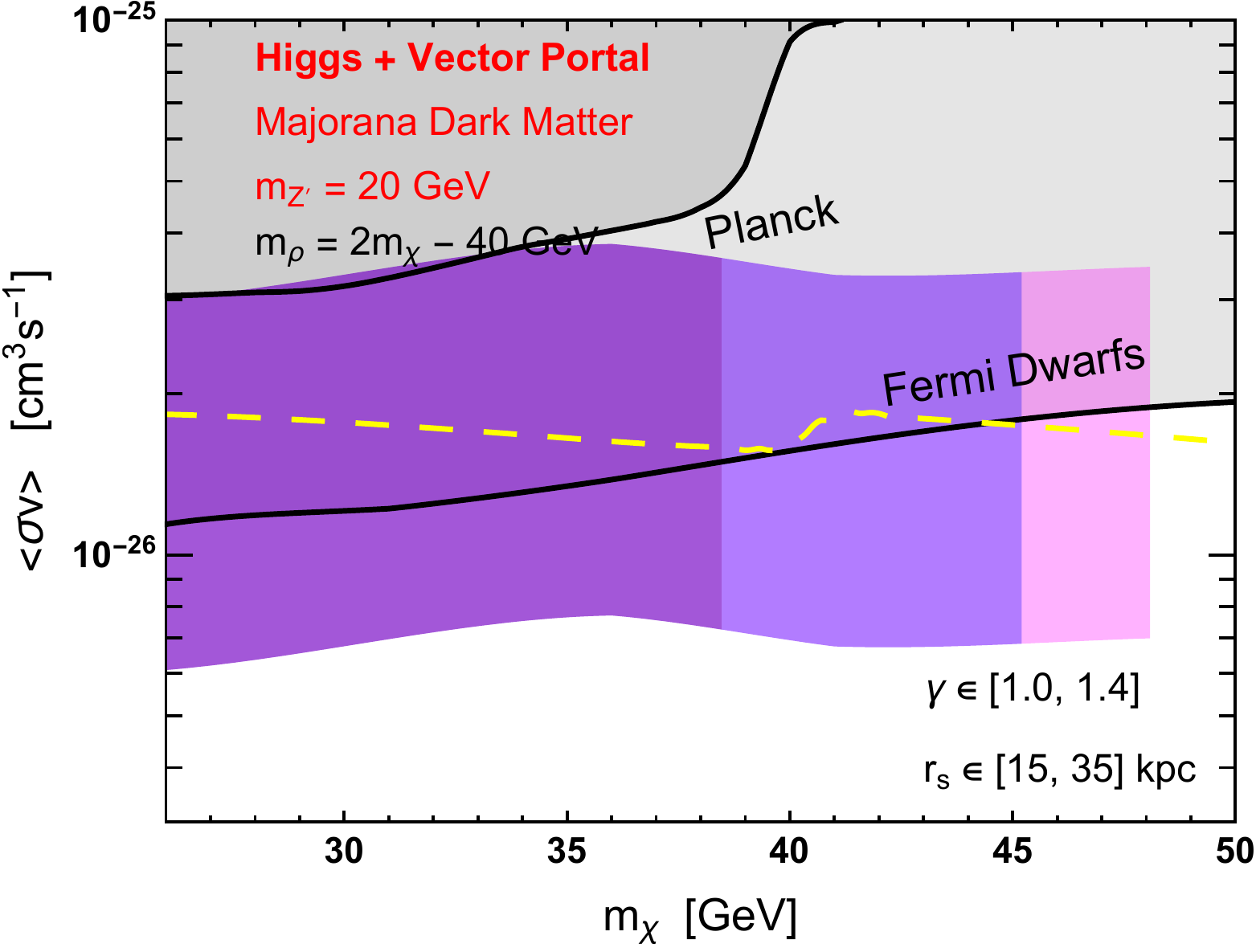}
\includegraphics[width=.49\textwidth]{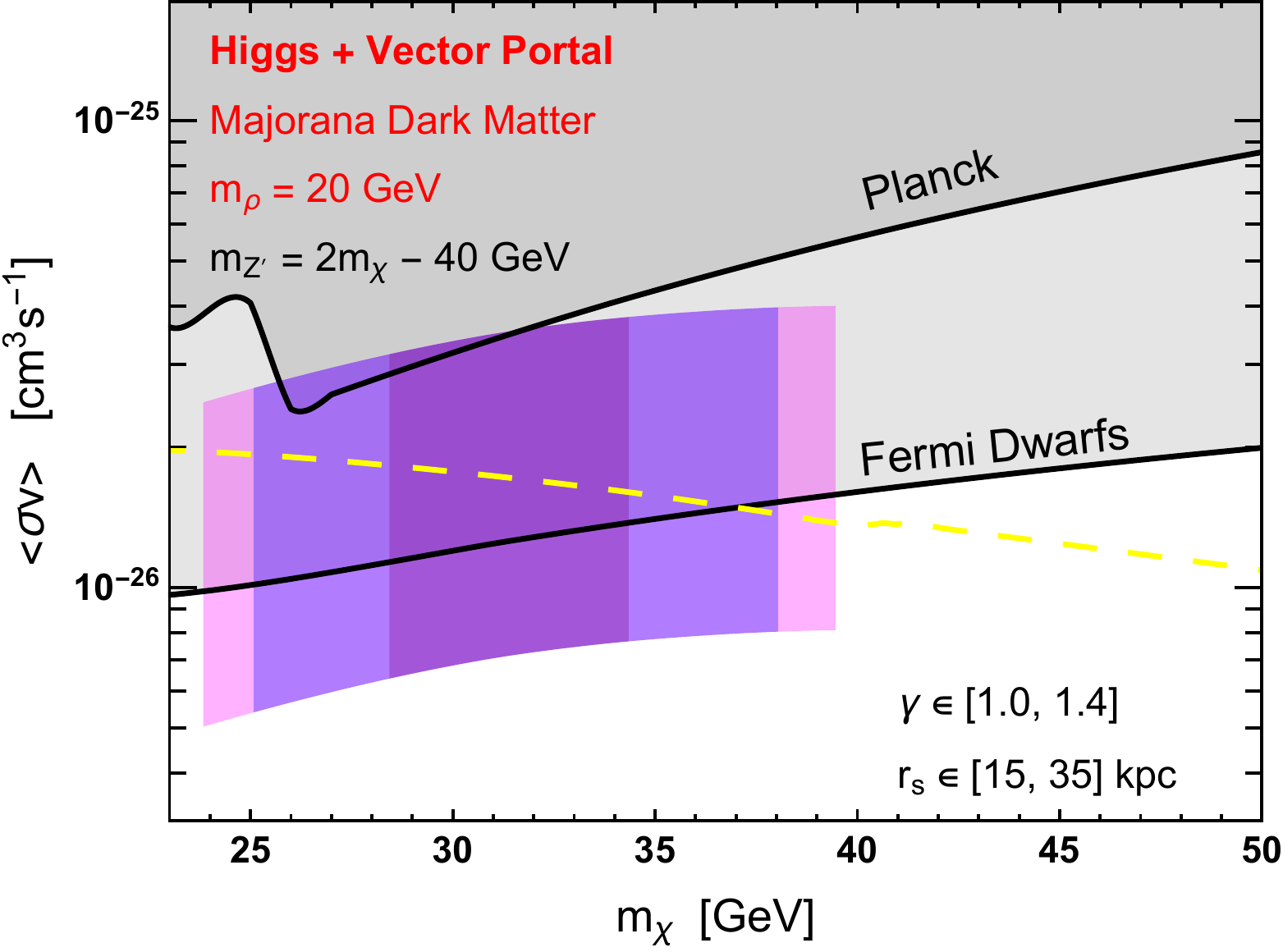}
\includegraphics[width=.49\textwidth]{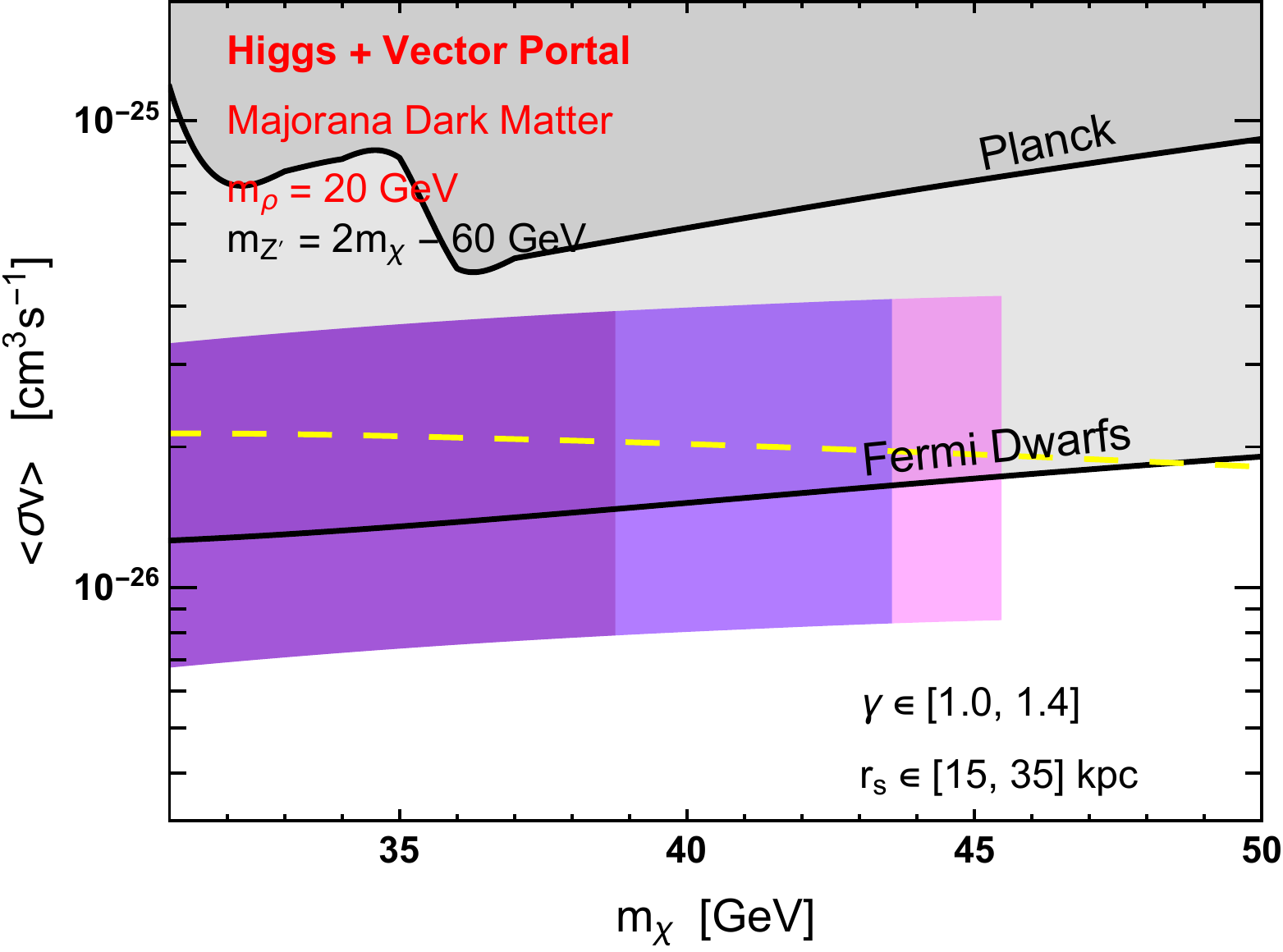}
\caption{\label{fig:higgsvector_param_space} As in Figs.~\ref{fig:photon_param_space} and~\ref{fig:higgs_param_space}, but for dark matter in the form of a Majorana fermion that annihilates to combinations of scalars and vectors which decay to the Standard Model through the Higgs and vector portals, respectively (see Sec.~\ref{subsec:Majorana}). This model can account for the entirety of the dark matter and generate the observed characteristics of the Galactic Center gamma-ray excess over substantial regions of parameter space.}
\end{figure*}

In Fig.~\ref{fig:higgs_param_space}, we repeat this exercise for the case of a vector dark matter candidate that annihilates to hidden sector scalars which then decay through the Higgs portal (see Sec.~\ref{subsec:Vector}). In this model, we find an even larger region of parameter space that can account for the entirety of the dark matter and generate the observed characteristics of the Galactic Center gamma-ray excess, corresponding to for $m_{X} \sim 70-110$ GeV and $m_{\rho} \gsim 10$ GeV.


Lastly, in Fig.~\ref{fig:higgsvector_param_space}, we show results for the case of Majorana dark matter which annihilates to combinations of scalars and vectors which decay to the Standard Model through the Higgs and vector portals, respectively. We show results for several selected values of $m_{\rho}$ and $m_{Z'}$. We find that this model can accommodate the observed features of gamma-ray excess with either $ m_{Z'} \lesssim 40\, \text{GeV}$ or $m_{\rho} \lesssim 70 \, \text{GeV} $.

\section{Additional Constraints}
\label{otherconstraints}

\begin{figure*}
\center
\includegraphics[width=.495\textwidth]{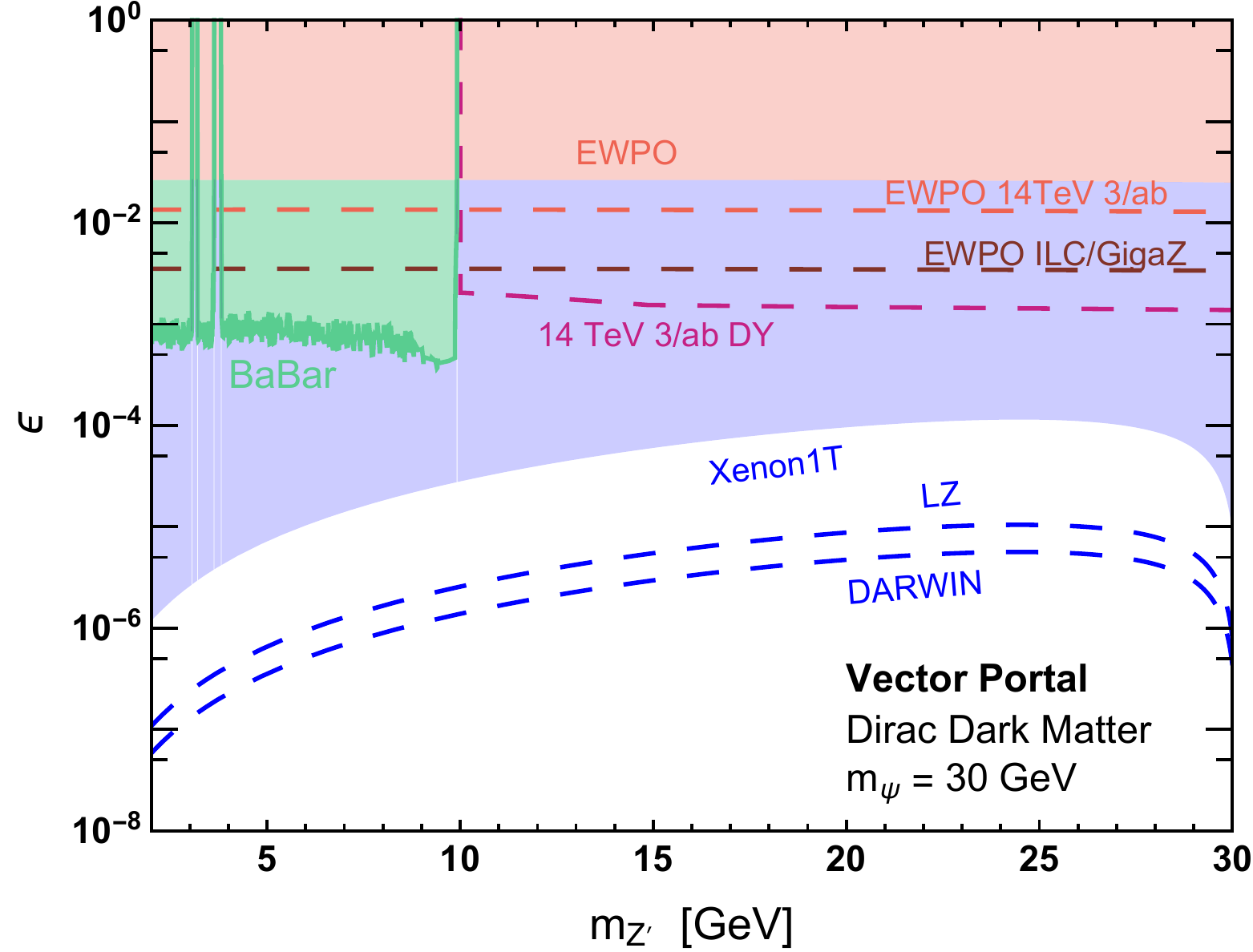}
\includegraphics[width=.495\textwidth]{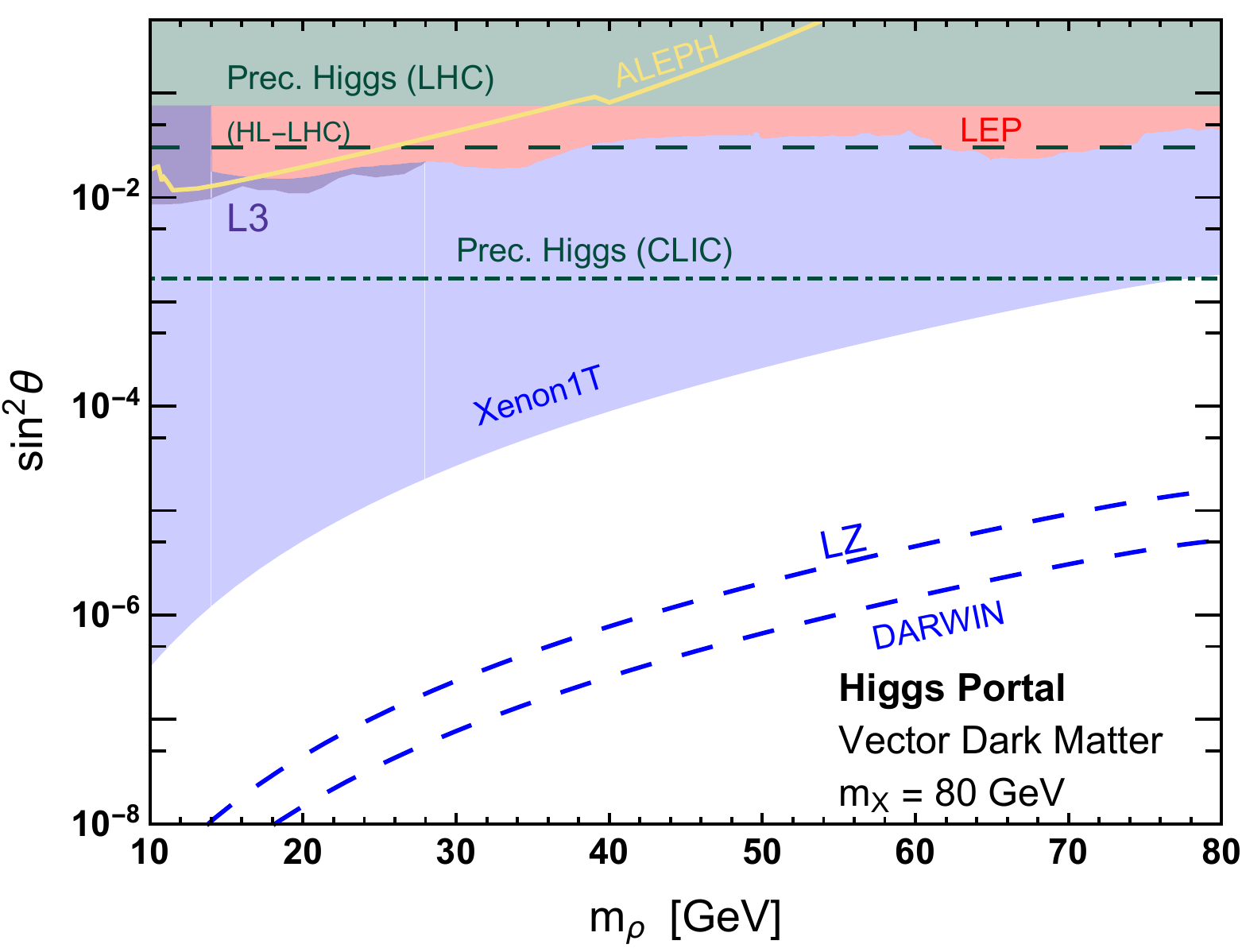}
\caption{ \label{fig:fermionDM_vectorMED_constraint} Collider and direct detection constraints on the Dirac dark matter vector portal model (left) and the vector dark matter Higgs portal model (right). Constraints denoted by dashed or dot-dashed lines represent projected sensitivities. Note that in the case of Majorana vector portal dark matter, constraints from electroweak precision observables are more restrictive than those derived from direct detection experiments.}
\end{figure*}

In addition to constraints derived from direct detection experiments, one can also consider collider signals of the vector and Higgs portal scenarios. Current and projected sensitivities to the vector portal model have been in calculated in Ref.~\cite{Curtin:2014cca}, and we present these in the left frame of Fig.~\ref{fig:fermionDM_vectorMED_constraint}. More specifically, we plot the current constraints from electroweak precision observables at the LHC (orange), the projected sensitivity for the high luminosity LHC assuming $\sqrt{s} = $14 TeV  and $3{\rm ab}^{-1}$, and the projected sensitivity for a future collider such as the ILIC or Giga$Z$. Since the $Z'$ in this model couples directly to quarks, it can be produced at the LHC via Drell-Yan production, and we plot the projected sensitivity to this signal for the high luminosity LHC, assuming $\sqrt{s} = $14 TeV and $3~{\rm ab}^{-1}$. Across the parameter space shown, however, we find that the constraints from direct dark matter searches can be considerably more restrictive, in particular in the case of Dirac dark matter. In addition to current constraints, we also present projected constraints from LUX-ZEPLIN (LZ) and DARWIN, as calculated and presented in Ref.~\cite{Witte:2017qsy}.

Light scalar particles which mix with the Standard Model Higgs boson have been probed extensively by LEP.  LEP probes the Higgs-scalar mixing angle thorough the indirect interaction of the hidden sector scalar and the Standard Model $Z$. This constraint is shown in the right frame of Fig.~\ref{fig:fermionDM_vectorMED_constraint} for a combination of data from LEP~\cite{Barate:2003sz}, ALPEH~\cite{Buskulic:1993gi} and L3~\cite{Acciarri:1996um}. 

Precision measurements of the Higgs couplings at the LHC can also be used to constrain the Higgs portal mixing angle. Here, we consider a number of searches performed by both ATLAS and CMS~\cite{Khachatryan:2016vau,Khachatryan:2016zqb,CMS-PAS-HIG-16-021,cms2017measurements,CMS:2017rli,CMS-PAS-HIG-17-010,ATLAS:2017myr,ATLAS:2016gld,ATLAS:2017cju,ATLAS:2016awy,ATLAS:2017bic} for various Higgs production channels and decay modes, performing a global fit of these measurements and presenting the resulting exclusion contours in the right frame of Fig.~\ref{fig:fermionDM_vectorMED_constraint}. We then repeated this exercise for the projected sensitivity of the high luminosity LHC~\cite{Slawinska:2016zeh,CMS:2017cwx} and the Compact Linear Collider (CLIC)~\cite{Abramowicz:2016zbo}. As shown in the right frame of Fig.~\ref{fig:fermionDM_vectorMED_constraint}, even these projected constraints are unlikely to be competitive with those derived from direct detection experiments. Again, these constraints are clearly much less restrictive than those derived from the results of direct detection experiments, with the exception of the case in which $m_\rho \sim m_h$.

\newpage

\section{Summary and Outlook}

In light of recent results from both direct dark matter searches and the LHC, dark matter models that are capable of generating the Galactic Center gamma-ray excess have become increasingly tightly constrained~\cite{Escudero:2016kpw}. It is straightforward to evade these constraints, however, within the context of models in which the dark matter does not directly couple to the Standard Model, but instead annihilates into unstable particles that reside within a hidden sector.  In this paper, we have revisited this class of models and demonstrated that they generically contain a broad range of parameter space that is capable of self-consistently generating the spectral shape and intensity of the observed gamma-ray excess.

\begin{figure*}
\center
\includegraphics[width=.48\textwidth]{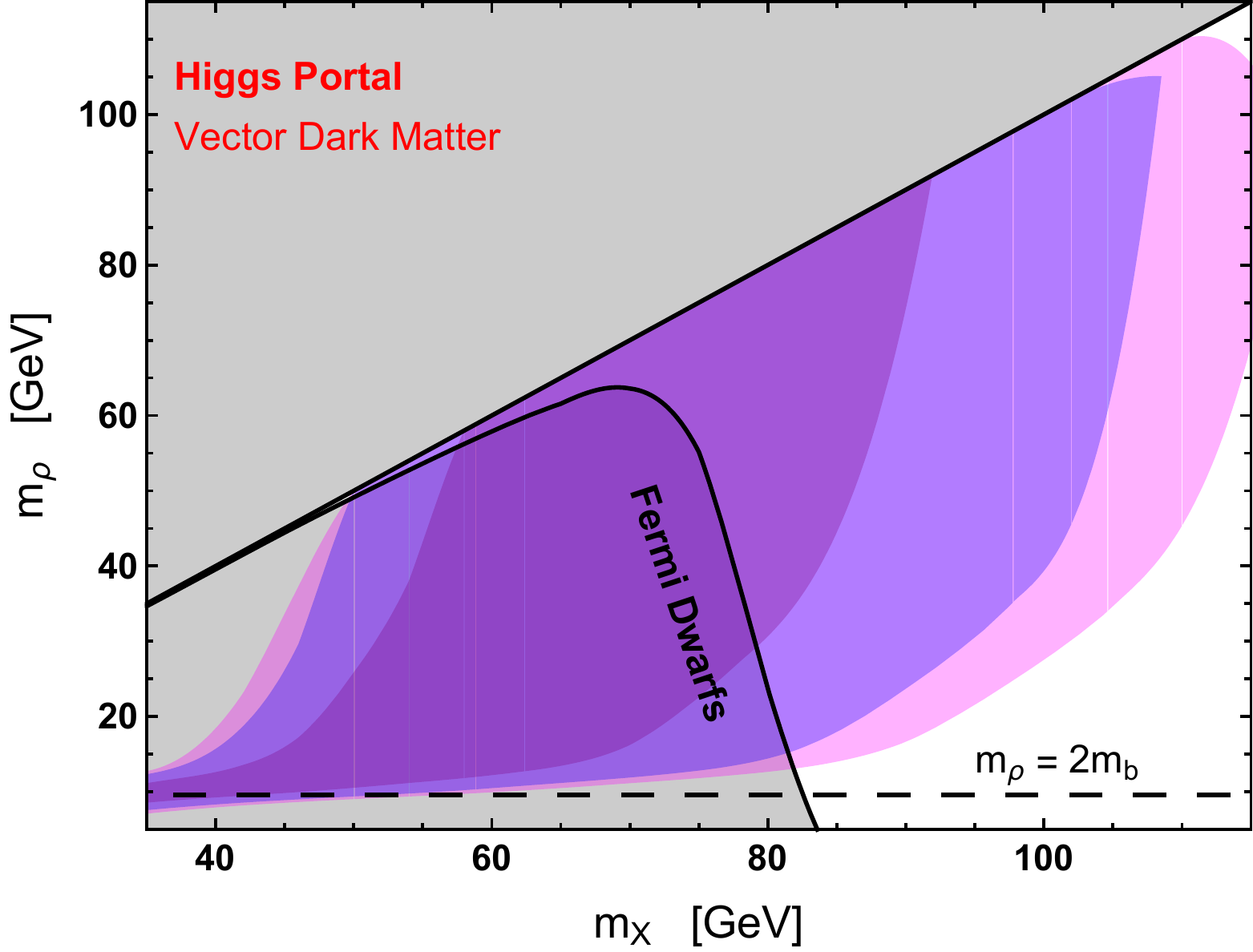} \\
\includegraphics[width=.48\textwidth]{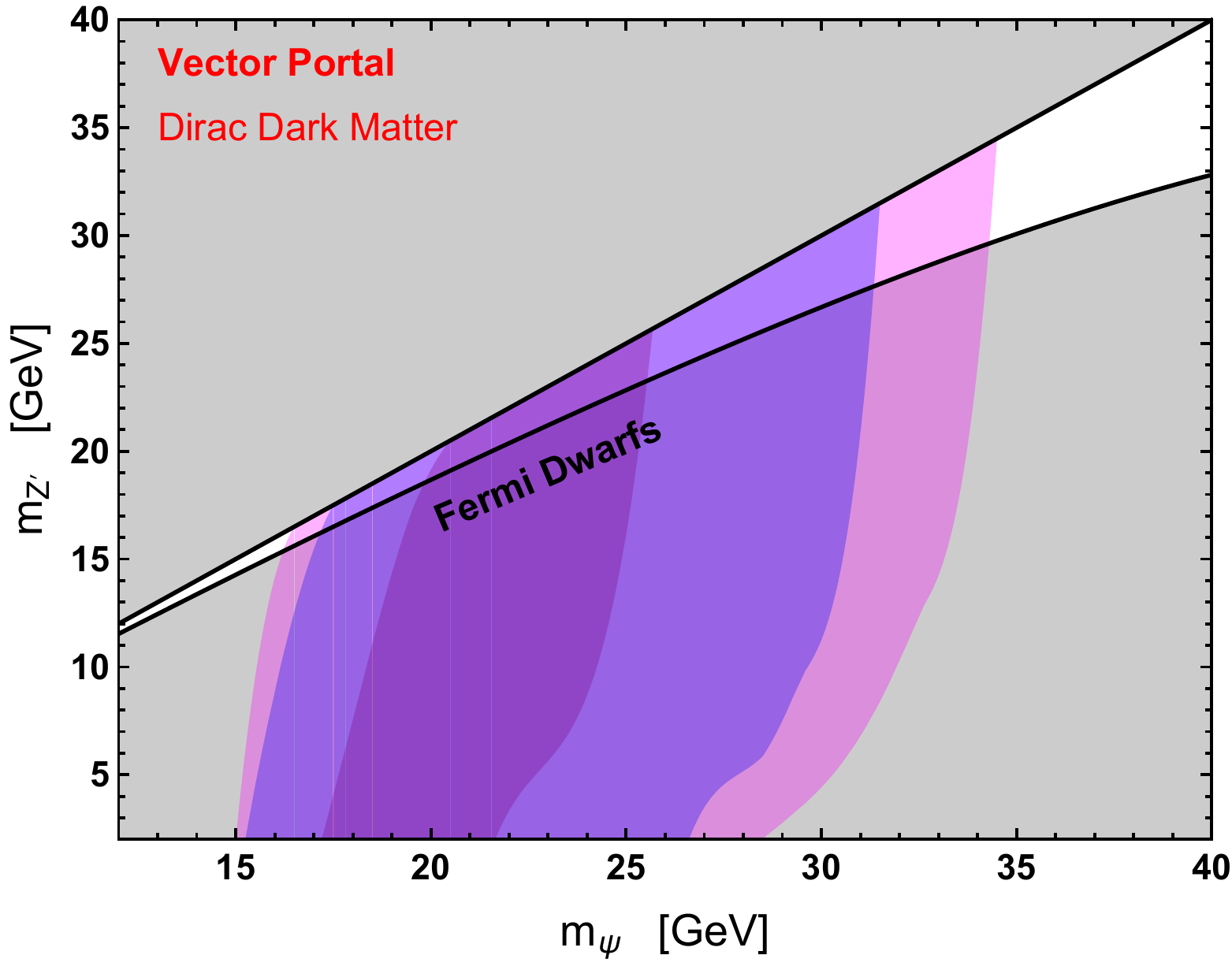}
\includegraphics[width=.48\textwidth]{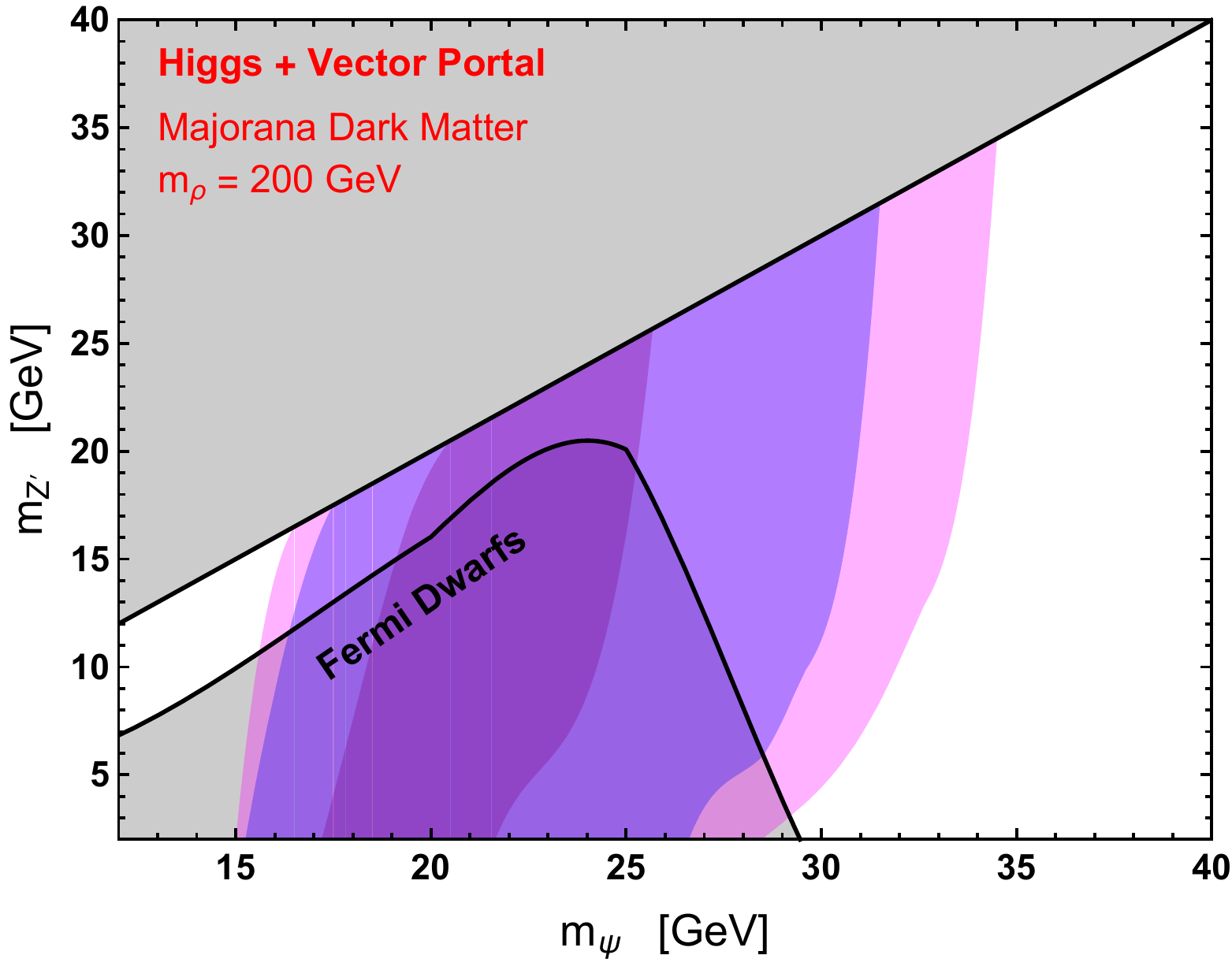} \\
\includegraphics[width=.48\textwidth]{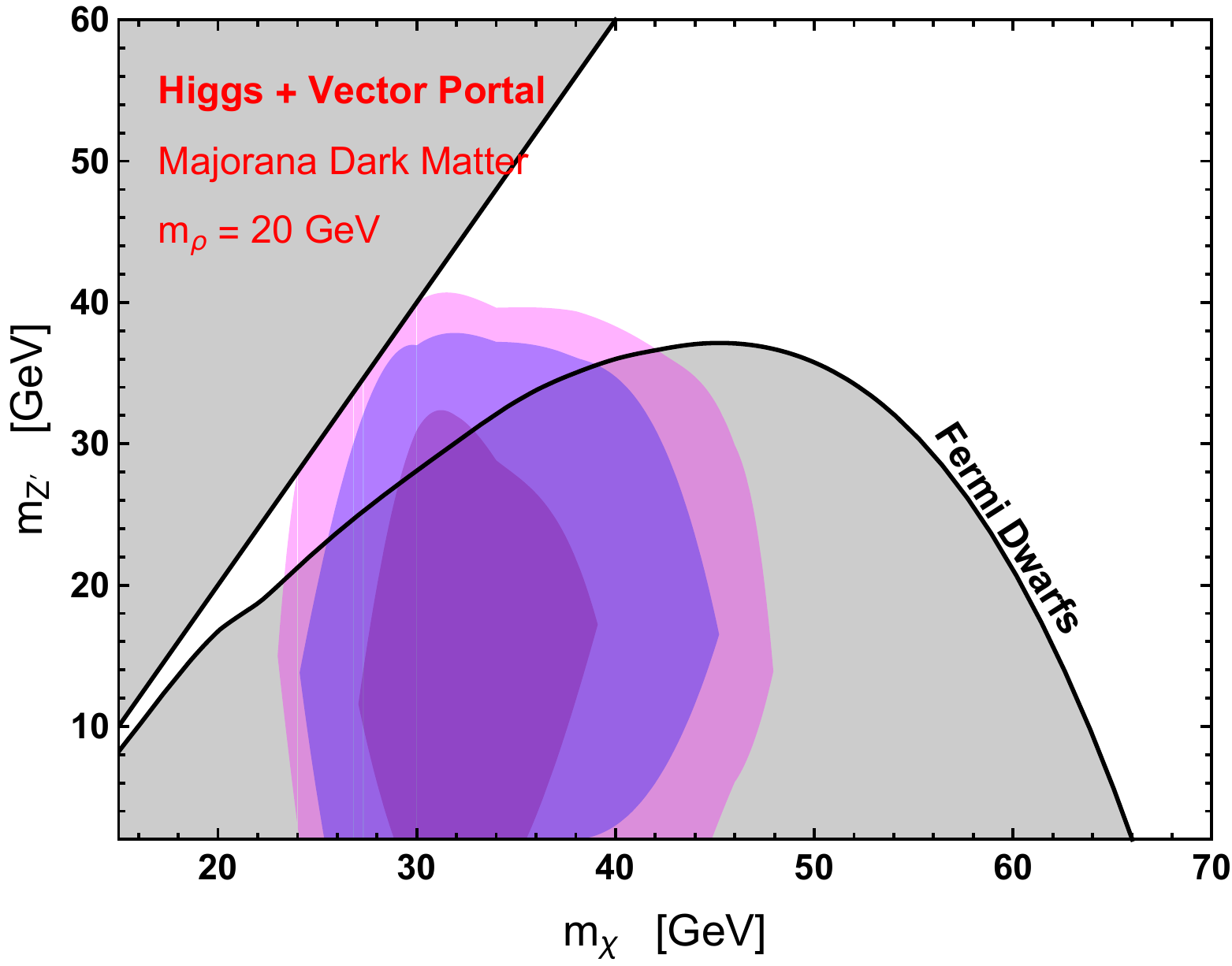}
\includegraphics[width=.48\textwidth]{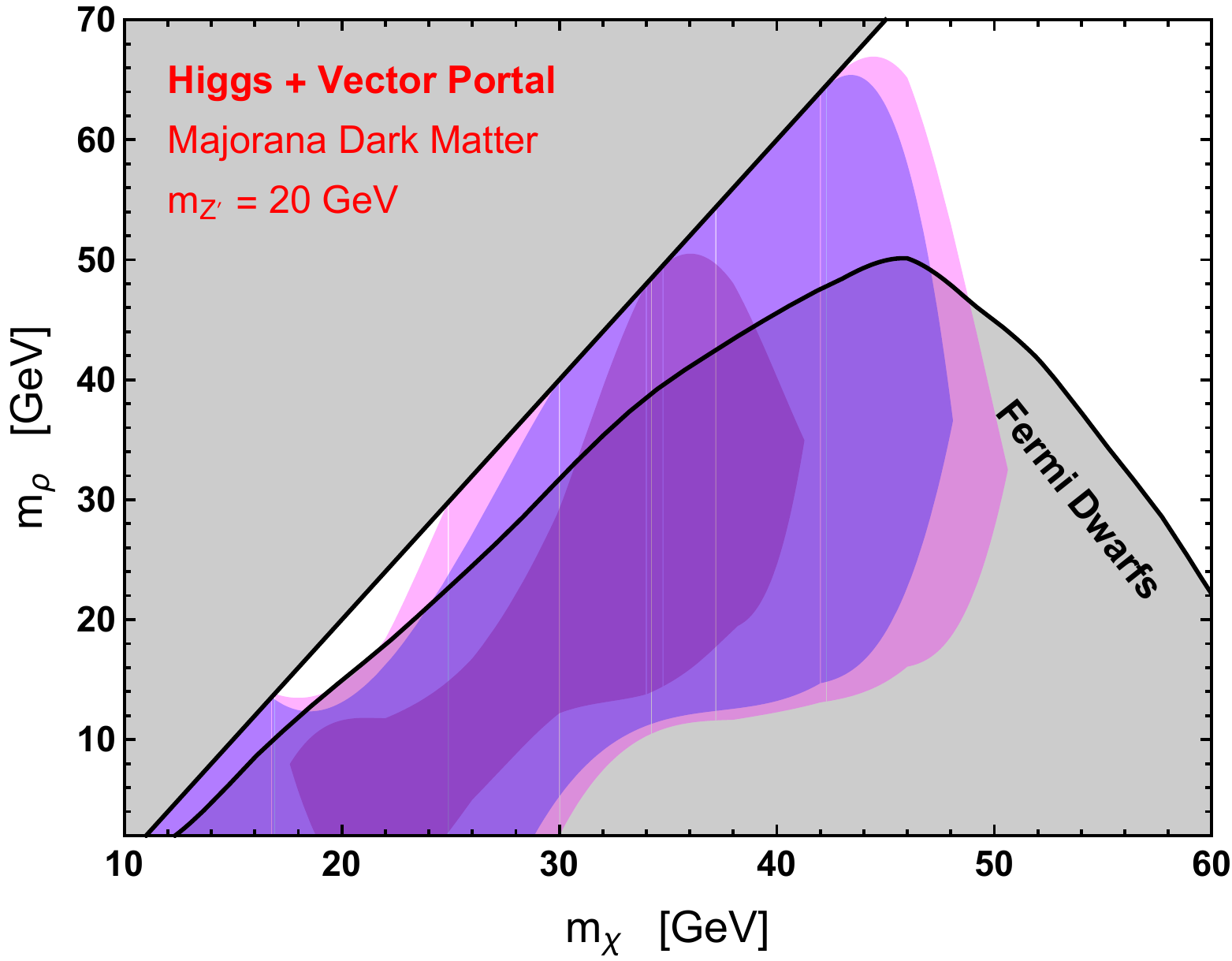}
\caption{ \label{fig:summary} A summary of the parameter space in the hidden dark matter models considered in this study. Throughout each frame, we have chosen the hidden sector coupling in order to obtain a thermal dark matter relic abundance equal to the measured cosmological dark matter density. The bands denote the regions in which the model provides a fit to the measured Galactic Center gamma-ray excess which yields a $p-$value of $\geq 0.32$ (dark purple), $\geq 0.10$ (violet), and $\geq 0.05$ (magenta), and we also show the regions that are disfavored by Fermi's observations of dwarf spheroidal galaxies~\cite{Fermi-LAT:2016uux}. In each of these models, there exists substantial parameter space that can accommodate the observed characteristics of the Galactic Center gamma-ray excess, while remaining consistent with other constraints.}
\end{figure*} 
 
Our main results are summarized in Fig.~\ref{fig:summary}. Throughout each frame, we have chosen the hidden sector coupling in order to obtain a thermal relic abundance equal to the measured cosmological dark matter density. The bands denote the regions in which the model provides a good fit to the measured Galactic Center gamma-ray excess (yielding a $p-$value greater than 0.32, 0.10, or 0.05), and we also show the regions that are disfavored by Fermi's observations of dwarf spheroidal galaxies~\cite{Fermi-LAT:2016uux}. From the four frames of this figure, it is clear that there exists substantial regions of parameter space in each of the models considered that can accommodate the observed characteristics of the Galactic Center gamma-ray excess, while remaining consistent with other constraints.

In the case of Dirac vector portal dark matter (middle left), the constraints from dwarf galaxies lead us to favor the regions of parameter space in which the dark matter is not much heavier than $m_{Z'}$. In contrast, Majorana dark matter is less restricted by dwarf constraints (middle right). We also find that the vector dark matter Higgs portal scenario (upper frame) is viable for $m_{X} \sim 70-110$ GeV and for a wide range of scalar masses. We find that Majorana dark matter with a combination of Higgs and Vector portals (lower frames) favors parameter space in which $m_{\chi} \sim 10-50$ GeV, $m_\rho \lsim 70$ GeV and $m_{Z'} \lsim 40$ GeV.

In many respects, the class of hidden sector models considered in this study is very difficult to test. Although future direct detection experiments will gradually become sensitive to hidden sectors that are even more decoupled from the Standard Model (\ie with even smaller values of $\epsilon$ or $\sin^2 \theta$ within the context of the vector portal and Higgs portal, respectively), viable parameter space will continue to exist well below the projected reach of such experiments. Similarly, the LHC will only be able to probe a relatively small fraction of the parameter space within this class of models. 
 
Unlike direct detection and collider experiments, however, dark matter annihilation signals are not generally suppressed in hidden sector models. In each of the models considered in this study, one predicts a gamma-ray flux from the Milky Way's population of dwarf spheroidal galaxies that likely to fall within the ultimate reach of the Fermi Gamma-Ray Space Telescope~\cite{Fermi-LAT:2016uux,Hooper:2015ula,Geringer-Sameth:2014qqa,Geringer-Sameth:2015lua,Li:2015kag}, after including the anticipated discoveries of new dwarf galaxies by DES and LSST~\cite{Charles:2016pgz}.

Measurements of the cosmic-ray antiproton spectrum by AMS-02 are also expected to be sensitive to much of the hidden sector dark matter parameter space that has been considered in this study. Intriguingly, a $\sim$4.5 $\sigma$ excess has been reported in this channel, peaking at energies of approximately $\sim$10-20 GeV~\cite{Cuoco:2016eej,Cui:2016ppb} (see also Ref.~\cite{Hooper:2014ysa}). Furthermore, the characteristics of the antiproton and gamma-ray excesses suggest that they could potentially be generated by annihilations of the same dark matter candidate.

\begin{figure}
\center
\includegraphics[width=.6\textwidth]{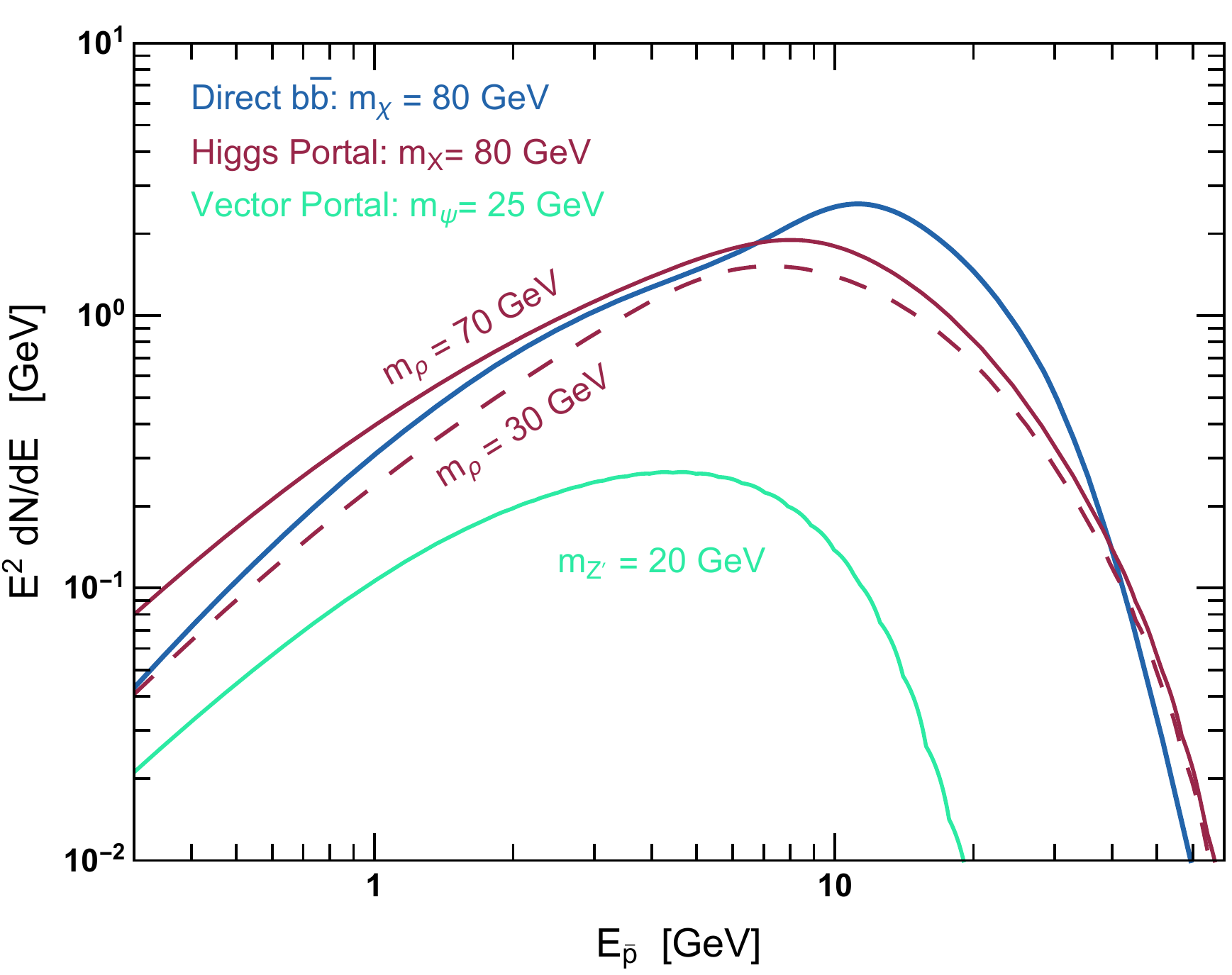}
\caption{ \label{fig:antiproton} The spectrum of antiprotons generated per dark matter annihilation, prior to any transport through the Galaxy, for several choices for the mass and annihilation channel. For the simple case of a dark matter candidate that annihilates directly to $b\bar{b}$, masses in the range of $\sim$50-90 GeV have been shown to be able to accommodate the spectral shape of the antiproton excess~\cite{Cuoco:2016eej,Cui:2016ppb}. The Higgs portal models shown predict an antiproton signal that would be very difficult to distinguish from that of a more conventional WIMP. In contrast, the antiproton flux is significantly suppressed in the vector portal model shown, especially at $E_{\bar{p}}\gsim 10$ GeV.}
\end{figure}

In Fig.~\ref{fig:antiproton}, we show the spectrum of antiprotons that is predicted to be generated per dark matter annihilation, prior to any transport through the Galaxy, for several choices for the dark matter's mass and annihilation channel. For the simple case of a dark matter candidate that annihilates directly to $b\bar{b}$, masses in the range of $\sim$50-90 GeV have been shown to be able to accommodate the spectral shape of the antiproton excess~\cite{Cuoco:2016eej,Cui:2016ppb}. In this figure, we compare this to the case of a Higgs portal model with $m_{X}=80$ GeV and $m_{\rho}=30$ GeV or 70 GeV, and for a vector portal model with $m_{\psi}=25$ GeV and $m_{Z'}=20$ GeV (each of which provide a good fit to the gamma-ray excess). The Higgs portal models shown predict an antiproton signal that would be very difficult to distinguish from that of a more conventional WIMP. In contrast, the suppression of the antiproton flux (especially at $E_{\bar{p}}\gsim 10$ GeV) in the vector portal model is rather distinctive, and could provide a way to discriminate this model from other dark matter scenarios that are capable of generating the observed gamma-ray excess.

\bigskip
\bigskip

\textbf{Acknowledgments:} ME would like to thank the Fermilab Theoretical Physics Department for
hospitality. ME is supported by Spanish Grant FPU13/0311 of MECD. ME also is supported by the European Union's Horizon 2020 research and innovation programme under the Marie Sk\l odowska-Curie grant agreement No
690575. SJW is supported by the European Union's Horizon 2020 research and innovation program under the Marie Sk\l odowska-Curie grant agreement No.~674896. This work was performed in part at Aspen Center for Physics, which is supported by National Science Foundation grant PHY-1607611. This manuscript has been authored in part by the Fermi Research Alliance, LLC under Contract No. DE-AC02-07CH11359 with the U.S. Department of Energy, Office of Science, Office of High Energy Physics. The United States Government retains and the publisher, by accepting the article for publication, acknowledges that the United States Government retains a non-exclusive, paid-up, irrevocable, world-wide license to publish or reproduce the published form of this manuscript, or allow others to do so, for United States Government purposes.

\bibliographystyle{JHEP}
\bibliography{gc_hidden}

\end{document}